\documentclass[11pt]{article}

\usepackage[utf8]{inputenc}
\usepackage[T1]{fontenc}
\usepackage{geometry}
\usepackage{amsmath}
\usepackage{amssymb}
\usepackage{amsthm}
\usepackage{mathtools}
\usepackage{graphicx}
\usepackage{rotating}
\usepackage{mathrsfs}
\usepackage{url}
\usepackage{xcolor}
\usepackage{tikz}
\usetikzlibrary{arrows.meta, positioning}
\usepackage{booktabs}
\usepackage{caption}
\usepackage{subcaption}
\usepackage{enumitem}
\usepackage{braket}
\usepackage[round]{natbib}
\usepackage{comment}
\usepackage{tcolorbox}
\usepackage{algorithm}
\usepackage{algorithmic}
\usepackage[hidelinks]{hyperref}
\usepackage{cleveref}

\newcommand{\cumulantGeneratingFunction}{\psi}
\newcommand{\naturalParameters}{\boldsymbol{\theta}}
\newcommand{\marginalEntropyLagrangeMultiplier}{\nu}
\newcommand{\gameTime}{\tau}
\newcommand{\variables}{\mathbf{x}}

\newcommand{\constraint}{\mathscr{C}}
\newcommand{\lagrangian}{\mathscr{L}}
\newcommand{\sufficientStatistics}{\mathbf{T}}
\newcommand{\sufficientStatisticMoments}{\boldsymbol{\mu}}

\ifdefined\draftmode
\usepackage{draftwatermark}
\SetWatermarkText{DRAFT}
\SetWatermarkScale{3}
\fi

\theoremstyle{definition}

\theoremstyle{remark}

\ifdefined\draftmode

\else
\excludecomment{draftcomment}
\fi

\begin{document}

\title{The Inaccessible Game}

\author{Neil D. Lawrence\\
Computer Laboratory\\
University of Cambridge\\
15 J. J. Thomson Avenue\\
Cambridge CB3 0FD, UK\\
\url{ndl21@cam.ac.uk}}

\date{10th November 2025}

\maketitle

\begin{abstract}
  In this paper we introduce the \emph{inaccessible game}, an information-theoretic dynamical system constructed from four axioms. The first three axioms are known and define \emph{information loss} in the system. The fourth is a novel \emph{information isolation} axiom that assumes our system is isolated from observation, making it observer-independent and exchangeable. Under this isolation axiom, total marginal entropy is conserved: $\sum_i h_i = C$. We consider maximum entropy production in the game and show that the dynamics exhibit a GENERIC-like structure combining reversible and irreversible components. 
\end{abstract}

\section{Introduction}

It is common to think of information as flowing, and information theory \citep{Shannon-mathematical48} gives us notions of information capacity. More recently, when comparing intelligences, \citet[\emph{The Atomic Human}][]{Lawrence-atomic24} focuses on constraints that dictate how information moves. The book uses the notion of an information topography ``In geography, the topography is
the configuration of natural and man-made features in the
landscape... These questions are framed by the topography. An
information topography is similar, but instead of the movement of
goods, water and people, it dictates the movement of information.''
However, no formal definition of an information topography is
given. This leaves the concept as a metaphor without mathematical
teeth. 

This paper is an attempt to address that weakness. In it we formalise a class of information dynamics. We do so by introducing the \emph{inaccessible game}, an abstract mathematical framework built on three established axioms and one novel postulate. The game builds on ideas from category theory that use three axioms to show that information loss is uniquely measured by (scaled) entropy difference
\citep{Baez-characterisation11}. We add a fourth postulate of information isolation which, under exchangeability and extensivity requirements, motivates conservation of total marginal entropy. The dynamics of the game emerge from constrained maximum entropy production within the information geometry. We show that these dynamics exhibit a GENERIC-like structure containing both dissipative and conservative elements and relate those dynamics to the physics of our natural world.

The inaccessible game follows in a tradition where simple rules can lead to complex emergent phenomena. John Conway's Game of Life
\citep{Gardner-life70} and Stephen Wolfram's one-dimensional cellular
automata \citep{Wolfram-cellular83} show that even when dynamics are
deterministic zero-player games can exhibit significant complexity that is
difficult to predict. \cite{Wolfram-undecidability85} called this behaviour computational irreducibility.

Our game differs from this tradition in that the rules
emerge from the \emph{information geometry} \citep{Amari-information00}. The dynamics are regulated through the Fisher information which operates as a conductance tensor and provides us with our information topography. 

The rest of this paper is organised as follows. We first review the information loss axioms of \citet{Baez-characterisation11} which give us our notion of information as represented by entropy (Section~\ref{sec-axiom-information-conservation}). We then introduce the fourth axiom, \emph{information isolation}, that constrains the system and makes the game inaccessible to an external observer. In Section~\ref{sec-dynamics-of-the-inaccessible-game} we introduce an information relaxation principle that governs the dynamics of the game through constrained maximisation of entropy production
\citep[MaxEnt,][]{Jaynes-information57,Jaynes-information63,Jaynes-minimum80,Ziegler-maximal87,Martyushev-maximum06,Beretta-nonlinear09} and show that these dynamics conform to the GENERIC structure, simulating them in Ising model variants in Section~\ref{sec-computational-demo} and \ref{sec-thermodynamic-limit}. We relate our entropy constraint to GENERIC's more traditional energy constraints in Section~\ref{sec-energy-entropy-equivalence}. Our final demonstration of the approach (Section~\ref{sec-mass-spring}) focusses on simple harmonic motion in a mass-spring system. 

\subsection*{Assumptions and Derivations}

To aid the reader, we clarify what is axiomatic, what is a modelling restriction, and what is derived.
\begin{itemize}[leftmargin=*]
\item \emph{Axiomatic:} Axioms 1--3 (functoriality, convex linearity, continuity) from \citet{Baez-characterisation11} that uniquely determine information loss as scaled Shannon entropy.
\item \emph{Postulated:} Axiom 4 (information isolation) stating the system is isolated from observation. Under exchangeability and extensivity requirements, this implies conservation of total marginal entropy: $\sum_i h_i = C$. Additionally, we postulate an \emph{information relaxation principle}: the system evolves to maximise entropy production subject to the marginal entropy constraint.
\item \emph{Modelling restriction:} We restrict to exponential family distributions parametrised by natural parameters $\naturalParameters$. This enables tractable computation but is not yet proven necessary from the axioms alone.
\item \emph{Derived:} The specific form of the constrained dynamics, the GENERIC-like decomposition, and the energy-entropy equivalence in the thermodynamic limit all follow from the above postulates and restrictions.
\end{itemize}

\section{Axioms and State Space}
\label{sec-axioms-and-state-space}

\citet{Shannon-mathematical48} introduced entropy as a measure of
information derived from probability through five axioms. More
recently \citet{Baez-characterisation11} showed that entropy can be
derived through category theory as a way of measuring
\emph{information loss} of a measure-preserving function. Without
invoking probability, they showed that the Shannon entropy emerges
from three axioms for information loss: functoriality, convex
linearity, and continuity. From these axioms they derive the information
loss as the difference between two Shannon entropies. Of particular interest to us is the second axiom, convex linearity.

\emph{Functoriality} suggests that given a process consisting of two stages, the amount of information lost in the whole process is the sum of the amounts lost at each stage
$$
F(f \circ g) = F(f) + F(g),
$$
where $\circ$ represents composition. \emph{Convex linearity} suggests that if we flip a probability-$\lambda$ coin to decide whether to do one process or another, the information lost is $\lambda$ times the information lost by the first process plus $(1-\lambda)$ times the information lost by the second,
$$
F(\lambda f \oplus (1-\lambda)g) = \lambda F(f) + (1-\lambda)F(g).
$$
The third axiom, \emph{continuity}, suggests that if we change a process slightly, the information lost changes only slightly, i.e. $F(f)$ is a continuous function of $f$.

The main result of their work is that 
\begin{quote}
  ... there exists a constant $c\geq 0$ such that for any $f: p \rightarrow q$
  $$
  F(f) = c(H(p) -H(q))
  $$
\end{quote}
where $F(f)$ is the information loss in process $f: p\rightarrow q$ and $H(\cdot)$ is the the Shannon entropy measured before and after the process is applied to the system.

\subsection{The Fourth Axiom: Information Isolation}
\label{sec-axiom-information-conservation}

For the inaccessible game, we introduce a fourth axiom. In an isolated chamber, mass and energy are conserved; our fourth axiom says that our game is similarly isolated from external observation. Under additional requirements of exchangeability and extensivity (see Appendix~\ref{app-information-isolation}), information isolation implies that the total marginal entropy should be conserved. 

For any finite sub-group of $N$ variables the
sum of marginal entropies, $\{h_i\}_{i=1}^{N}$, sums to a constant, $C$,
\begin{equation}
\sum_{i=1}^N h_i = C,
\label{eq-marginal-entropy-constraint}
\end{equation}
or in other words the total amount of information in the system is
conserved. The information conservation constraint is imposed in an
exchangeable form across the \emph{marginal} entropies. The form
ensures that we can consider any finite partition of variables from a
total that \emph{could} be countably infinite. The exchangeability
idea is inspired by, but differs from, the more usual use of
exchangeability in non-parametric probability
\citep{deFinetti-funzione31,Aldous-exchangeability85,Bernardo-bayesian00}.

\citeauthor{Baez-characterisation11}'s three axioms provide the
foundational notion of entropy loss that justifies our use of marginal
entropy in our information conservation axiom. 

In traditional thermodynamics, energy conservation plays the role of an isolated-system constraint. Here, marginal entropy conservation plays an analogous role: it defines a built-in potential within the information geometry. As we show in Section~\ref{sec-dynamics-of-the-inaccessible-game}, the curvature of this potential, encoded in the Fisher information, acts as an intrinsic metric governing how the system redistributes its informational energy content.

\section{Dynamics of the Inaccessible Game}
\label{sec-dynamics-of-the-inaccessible-game}

In this section we introduce an \emph{information relaxation} principle, to connect the conservation law to the dynamics of the game. The principle suggests that the system evolves to maximise entropy production while being subject to the conservation constraint. We enforce $\sum_i h_i = C$ (Axiom 4), and choose dynamics that maximise $\dot{H}$ subject to this. Equivalently, the game relaxes multi-information while keeping total marginal entropy fixed. This constrained maximum entropy production leads to dynamics that combine dissipative and conservative components.

The multi-information (also called total correlation), $I$, is a measure of correlation \citep{Watanabe-multiinformation60}. It is given by sum of marginal entropies minus the joint entropy, $H$,
$$
I = \sum_{i=1}^n h_i - H.
$$
Because the joint entropy is always less than or equal to the sum of marginals, multi-information is always non-negative. It is zero when the variables are fully independent and positive if there is any correlation between them.

We can use the definition of multi-information to rewrite the marginal entropy constraint $\sum_{i=1}^n h_i = C$ using the multi-information $I$ and joint entropy $H$,
\begin{equation}
I + H = C,
\label{eq-information-relaxation-principle}
\end{equation}
where the joint entropy is maximised when the system is totally independent and the multi-information is maximised when the system is fully correlated.

This identity does \emph{not} by itself impose dynamics; it only re-expresses the conservation constraint. The dynamics are obtained by maximising entropy production of $H$ subject to the constraint $\sum_i h_i = C$. Since $I + H = C$, maximising $H$ is equivalent to minimising $I$, so we can view the system as relaxing multi-information while staying on the constraint surface. The constraint gradient $\nabla(\sum_i h_i)$ enforces tangency to the surface. In Section~\ref{sec-energy-entropy-equivalence} we will show how this direction aligns with constraint gradients from more traditional energy, $\nabla E$, in the thermodynamic limit. The game starts in the fully correlated state and evolves to maximise joint entropy $H$ while preserving the marginal entropy constraint. 

These relaxation dynamics require us to parametrise the entropy. Entropy is a functional that is dependent on probability distributions. We will restrict ourselves\footnote{We will attempt to justify this restriction in later work, for now though you may consider it a limitation of our framework.} to considering exponential family distributions which are parametrised through natural parameters, $\naturalParameters$. 

For variables $\variables$ the exponential family is written in the form
$$
p(\variables) = \exp(\naturalParameters^{\top} \sufficientStatistics(\variables) - \cumulantGeneratingFunction(\naturalParameters)),
$$ 
where the natural parameters are the coefficients of the sufficient statistics $\sufficientStatistics(\variables)$ and the log partition function $\cumulantGeneratingFunction(\naturalParameters)$ ensures normalisation. The log partition function is also the cumulant generating function of the distribution.\footnote{The derivatives of the cumulant generating function with respect to the natural parameters return the different cumulants of the distribution. The first derivative returns the mean, the second the covariance, then the skew etc..} 

Many characteristics of the exponential family are captured through the Fisher information matrix, 
$$
G(\naturalParameters) = \nabla^2 \cumulantGeneratingFunction(\naturalParameters),
$$
which is the Hessian of the log partition function. Since the log partition function is also the cumulant generating function, the Fisher information is also the \emph{covariance} of the sufficient statistics, $\sufficientStatistics(\variables)$ under $p(\variables)$. The joint entropy of an exponential family distribution also has a relatively straightforward form, it's written as
$$
H = \cumulantGeneratingFunction(\naturalParameters) - \naturalParameters^{\top} \sufficientStatisticMoments(\naturalParameters),
$$
where $\sufficientStatisticMoments(\naturalParameters)=\nabla \cumulantGeneratingFunction(\naturalParameters)$ is the mean of $\sufficientStatistics(\variables)$. This means the gradient of the joint entropy with respect to the natural parameters can be computed as,
$$
\nabla H = -G(\naturalParameters)\naturalParameters,
$$
where this simple form arises because terms that involve the first derivative of the cumulant generating function, $\nabla\cumulantGeneratingFunction(\naturalParameters)$, cancel. 

Note that unconstrained dynamics $\dot{\naturalParameters} = -G(\naturalParameters)\naturalParameters = \nabla H$ would constitute pure maximum entropy production (MEP) for $H$.\footnote{Here $\dot{\naturalParameters} = \tfrac{\text{d}\naturalParameters}{\text{d}\gameTime}$ is how the parameters evolve across the progression of the game as measured by $\gameTime$, which we call ``game time''.} This is gradient ascent on $H$ in the natural parameter coordinates. This form depends on our choice to use natural parameters; in other coordinate systems the dynamics would differ. The marginal entropy constraint $\sum_i h_i = C$ imposes an additional requirement: we must project the MEP flow onto the tangent space of the constraint manifold. The constraint will introduce an extra term $\marginalEntropyLagrangeMultiplier(\gameTime)\mathbf{a}(\naturalParameters)$ in equation \eqref{eq-constrained-dynamics}, where $\mathbf{a} = \nabla(\sum_i h_i)$ is the constraint gradient. Since $\sum_i h_i = I + H$, we have $\mathbf{a} = \nabla I + \nabla H$, so the multi-information gradient $\nabla I$ influences the dynamics through the constraint enforcement mechanism.

To ensure marginal entropy remains conserved we introduce a Lagrangian formulation,
$$
\lagrangian(\naturalParameters, \marginalEntropyLagrangeMultiplier) = -H(\naturalParameters) + \marginalEntropyLagrangeMultiplier\left(\sum_{i=1}^n h_i -C\right).
$$
where $\marginalEntropyLagrangeMultiplier$ is a Lagrange multiplier. Traditionally Lagrangians are minimised, so we've based ours around the negative joint entropy. This is in-line with our principle of relaxation of the multi-information. 

The information relaxation dynamics are now given by
\begin{equation}
\dot{\naturalParameters} = -G(\naturalParameters)\naturalParameters + \marginalEntropyLagrangeMultiplier(\gameTime) a(\naturalParameters),
\label{eq-constrained-dynamics}
\end{equation}
where $a(\naturalParameters) = \sum_i \nabla h_i(\naturalParameters)$ and $\marginalEntropyLagrangeMultiplier(\gameTime)$ is determined by the constraint enforcement condition. Here we have parametrised the trajectory through the parameter space by $\gameTime$, which we call game time. This parameter measures progress along the path $\naturalParameters(\gameTime)$ traced out by the dynamics. 

The Fisher information $G(\naturalParameters)$ provides the natural Riemannian metric on the exponential family manifold. The term $-G(\naturalParameters)\naturalParameters = \nabla H$ represents gradient ascent on joint entropy in this geometry. Without the constraint, the system would follow this natural gradient flow, maximising $H$ along the steepest path defined by the information metric. However, the marginal entropy constraint $\sum_i h_i = C$ forces the dynamics away from this natural path. The Lagrange multiplier term $\marginalEntropyLagrangeMultiplier(\gameTime)\mathbf{a}(\naturalParameters)$ projects the flow onto the constraint surface, creating a balance between following the information geometry and maintaining the conservation law. This tension between the geometric structure and the constraint is what generates the GENERIC-like decomposition we analyse in the next section.

Since the constraint must be satisfied at all game times, $\gameTime$, we have
\begin{equation}
\frac{\text{d}}{\text{d}\gameTime}\left(\sum_i h_i\right) = 0.
\end{equation}
The $h_i$ values are derived from the underlying exponential family distributions. They are often difficult to compute as they involve marginalisation of $n-1$ variables from $p(\variables)$ to extract $p(x_i)$. This means that each $h_i$ is in general a function of all $\naturalParameters$. 

The system dynamics are derived by requiring that the constraint $\sum_i h_i = C$ is maintained exactly at all times. We can use the chain rule to enforce this constraint,
\begin{equation}
\mathbf{a}(\naturalParameters)^\top \dot{\naturalParameters} = 0.
\label{eq-constraint-maintenance}
\end{equation}
where 
$$
\mathbf{a}(\naturalParameters) = \nabla \sum_i h_i.
$$
Substituting in our constrained dynamics (equation \eqref{eq-constrained-dynamics}) we obtain
$$
\mathbf{a}(\naturalParameters)^\top \left(-G(\naturalParameters)\naturalParameters + \marginalEntropyLagrangeMultiplier(\gameTime) \mathbf{a}(\naturalParameters)\right) = 0
$$
and this allows us to solve for $\marginalEntropyLagrangeMultiplier(\gameTime)$ giving
\begin{equation}
  \marginalEntropyLagrangeMultiplier(\gameTime) = \frac{\mathbf{a}(\naturalParameters)^\top G(\naturalParameters)\naturalParameters}{\|\mathbf{a}(\naturalParameters)\|^2},
\label{eq-nu-solution}
\end{equation}
which can be substituted back in to the dynamics to give
$$
\dot{\naturalParameters} = -\Pi_\parallel(\naturalParameters) G(\naturalParameters)\naturalParameters,
$$
where
$$
\Pi_\parallel(\naturalParameters) = \mathbf{I} - \frac{\mathbf{a}(\naturalParameters)\mathbf{a}(\naturalParameters)^\top}{\|\mathbf{a}(\naturalParameters)\|^2}
$$
is a projection matrix that ensures the trajectory stays on the manifold defined by the constraint; the subscript $\parallel$ indicates projection onto the tangent space. Note that this is the \emph{Euclidean} projection onto the tangent space (parallel to the constraint manifold, orthogonal to $\mathbf{a}$ in the ambient parameter space), not projection in the Fisher metric, $G(\naturalParameters)$. 

The entropy of an exponential family has a unique global maximum at $\naturalParameters=\mathbf{0}$, i.e. at the reference measure with no bias from sufficient statistics. This means that our system will never be at equilibrium until it reaches its end-point. Instead of understanding equilibrium behaviour we look to understand the transient dynamics as the system moves along the constraint manifold prescribed by $\sum_i h_i = C$.

The Lagrange multiplier $\marginalEntropyLagrangeMultiplier(\gameTime)$ is given by a scaled inner product between the steepest ascent dynamics and the gradient of the constraint \eqref{eq-nu-solution}. It varies with the system state. In the next section we show how this variation drives a balance between entropy production and constraint enforcement forces. The character of the dynamics, whether more thermodynamic or more mechanical, depends on this balance. We show that the characteristics of these dynamics provide a GENERIC-like formalism. GENERIC (General Equation for Non-Equilibrium Reversible-Irreversible Coupling) is a framework from physics that characterises dissipative-conservative non-equilibrium systems \citep{Grmela-dynamics97,Ottinger-beyond05}. In what follows, to simplify notation, we will often drop the dependence of $G$, $\mathbf{a}$ and $\Pi_\parallel$ on $\naturalParameters$.

\section{A GENERIC-like Structure}
\label{sec-generic-structure}

In this section, we show that the constrained maximum entropy production gives rise to a GENERIC-like decomposition, combining dissipative and conservative components. GENERIC was developed by \citet{Grmela-dynamics97,Ottinger-beyond05} to describe systems that combine reversible and irreversible dynamics. It provides a unified mathematical structure for non-equilibrium thermodynamics that generalises both classical mechanics and thermodynamics. For our framework, the thermodynamic consistency conditions, that are typically imposed by hand in GENERIC, emerge automatically from the marginal entropy conservation constraint.

A GENERIC decomposition involves a \emph{symmetric} part that represents dissipative dynamics, i.e.\ entropy production that drives the system toward higher joint entropy. And an \emph{antisymmetric} part represents conservative dynamics, i.e.\ rotations on the constraint manifold that redistribute information changing the marginal entropy. The asymmetric component emerges from the geometric structure of the constraint. 

The relative strength of these dissipative and conservative components is determined by the Lagrange multiplier $\marginalEntropyLagrangeMultiplier(\gameTime)$. As $\marginalEntropyLagrangeMultiplier \to 0$, the constraint gradient operates orthogonally to the direction of the flow and dissipation dominates. As $\marginalEntropyLagrangeMultiplier$ increases the conservative part grows and the system transitions toward mechanical regimes with reduced entropy production.

The decomposition arises from the interplay between the information geometry and the constraint structure. The symmetric part combines contributions from the Fisher information $G$ (which governs the natural gradient flow on the manifold), third-order cumulant statistics (encoded in $\nabla G$), and the constraint Hessian. The antisymmetric part emerges specifically from the coupling between the constraint gradient $\mathbf{a}$ and the Lagrange multiplier gradient $\nabla\marginalEntropyLagrangeMultiplier$, creating an outer product $\mathbf{a}(\nabla\marginalEntropyLagrangeMultiplier)^\top$ whose antisymmetric component generates rotations on the constraint manifold. The full derivation is provided in Appendix~\ref{app-generic-sufficiency}.

In the standard GENERIC formulation, the dynamics of a state $\naturalParameters$ are written as
\begin{equation}
  \dot{\naturalParameters} =
  A(\naturalParameters)\nabla E(\naturalParameters) +
  S(\naturalParameters)\nabla H(\naturalParameters),
\label{eq-generic-standard}
\end{equation}
where $E$ is an energy functional, $H$ is an entropy functional, $A$ is an antisymmetric Poisson operator encoding reversible dynamics, and $S$ is a symmetric positive-semidefinite friction operator encoding irreversible dynamics.

The GENERIC framework requires two degeneracy conditions to ensure thermodynamic consistency. Firstly, the entropy should be conserved by the reversible dynamics,
\begin{equation}
  A(\naturalParameters)\nabla H(\naturalParameters) = 0.
\label{eq-degeneracy-1}
\end{equation}
Secondly, the energy should be conserved by the irreversible dynamics,
\begin{equation}
  S(\naturalParameters)\nabla E(\naturalParameters) = 0.
\label{eq-degeneracy-2}
\end{equation}
These degeneracy conditions are typically difficult to satisfy. In most applications, the operators $A$ and $S$ must be carefully constructed to ensure the degeneracies hold. Manual construction can be challenging (see Section~\ref{sec-mass-spring}).

Conversely, in the information dynamics we have derived, our degeneracy conditions emerge \emph{automatically}. The first condition, $A\nabla H = 0$, requires that the antisymmetric part of the dynamics conserves entropy. In our framework, this holds by construction at every point on the constraint manifold. It comes from the tangency condition from Section~\ref{sec-dynamics-of-the-inaccessible-game}. The constraint maintenance requirement,
\begin{equation}
  \mathbf{a}^\top \dot{\naturalParameters} = 0,
\end{equation}
where $\mathbf{a} = \nabla \left(\sum_i h_i \right)$ is the constraint gradient, ensures that the dynamics remain tangent to the constraint surface at all times. This tangency is not only a local property, it holds globally across the entire constraint manifold because it is enforced through the Lagrange multiplier $\marginalEntropyLagrangeMultiplier(\gameTime)$.

The antisymmetric part of the dynamics inherits this global tangency. Since the antisymmetric component generates entropy-conserving rotations (by definition, $\mathbf{z}^\top A \mathbf{z} = 0$ for antisymmetric $A$), and since these rotations must remain tangent to the constraint surface, the first degeneracy condition is automatically satisfied everywhere.

It's in the second degeneracy condition where our framework departs from the standard formulation of GENERIC which that the symmetric part conserves energy, $S\nabla E = 0$. This typically must be verified case-by-case. In our framework, the marginal entropy constraint $\sum_i h_i = C$ plays the role that energy conservation plays in GENERIC: its gradient defines the degeneracy direction along which the dissipative operator vanishes,
$$
S \nabla \left( \sum_i h_i \right)  = 0 \equiv S \mathbf{a} = 0.
$$
This degeneracy condition follows from the constraint tangency requirement. In Section~\ref{sec-energy-entropy-equivalence} we argue that, under certain conditions in the thermodynamic limit, this constraint gradient becomes asymptotically parallel to $\nabla E$, establishing a connection between the information-theoretic and classical physical pictures.

For GENERIC, constructing operators $A$ and $S$ that satisfy both degeneracy conditions typically requires significant effort \citep[see e.g.][Chapter 4, particularly Sections 4.2.2--4.2.3]{Ottinger-beyond05}. In our framework, these degeneracy-like properties emerge from the linearised flow structure, although we note that whether our form leads to a full global Poisson structure satisfying the Jacobi identity remains an open question.\footnote{Numerical investigations reveal that the Jacobi identity exhibits a dependence on parameter symmetry. For the harmonic oscillator, it holds to machine precision ($\sim 10^{-12}$) when $\theta_{xx} = \theta_{pp}$ (isotropic case) but fails with violations of order $10^{-3}$ for $\theta_{xx} \neq \theta_{pp}$, affecting 6 of 27 triplets. Symbolic verification using SymPy confirms these violations are genuine. Similar behavior emerges in N=3 binary variables: symmetric configurations satisfy the identity ($\sim 10^{-7}$) while asymmetric ones exhibit violations ($\sim 10^{-5}$). When scaled by $\|A\|_F^2$, both systems show comparable relative violations of order $10^{-1}$, suggesting a universal pattern. This connects parameter symmetry to rotational invariance in phase or configuration space, i.e.\ precisely the structure Poisson brackets capture. The isotropic harmonic oscillator admits SO(2) symmetry; permutation-symmetric spin systems admit discrete rotation symmetry. Breaking these symmetries destroys global Poisson structure, giving further evidence why deriving GENERIC \emph{ab initio} is notoriously difficult: most physical systems lack the required symmetries. See Appendix~\ref{app-generic-sufficiency} and supplementary verification code.} The fact that GENERIC-like properties emerge naturally suggests that marginal entropy conservation has special geometric significance.

To understand the local structure of the dynamics, we apply flow perturbation analysis \citep{Guckenheimer-dynamics83,Wiggins-introduction90,Strogatz-nonlinear94}. This reveals how the GENERIC-like decomposition emerges from the constraint geometry.

\subsection{Local Flow Analysis}
\label{sec-local-flow-analysis}

As we've highlighted,  entropy maximisation has a unique global maximum at $\naturalParameters=\mathbf{0}$, which the system will not reach during finite-time evolution. Therefore, rather than studying equilibrium behaviour, we study the \emph{transient dynamics} along the constraint manifold.

The constrained dynamics are given by
\begin{equation}
 \dot{\naturalParameters} = -G(\naturalParameters)\naturalParameters +
 \marginalEntropyLagrangeMultiplier(\naturalParameters) \mathbf{a}(\naturalParameters) \coloneqq
 F(\naturalParameters).
\label{eq-constrained-flow}
\end{equation}
For any point $\naturalParameters^\ast$ on the constraint manifold, define the displacement $\mathbf{q} = \naturalParameters - \naturalParameters^\ast$. The linearised flow near $\naturalParameters^\ast$ is
\begin{equation}
 \dot{\mathbf{q}} = M \mathbf{q} + O(\|\mathbf{q}\|^2),
\label{eq-linearised-flow}
\end{equation}
where $M = \tfrac{\partial F}{\partial\naturalParameters}\big|_{\naturalParameters^\ast}$ is the Jacobian of the flow evaluated at $\naturalParameters^\ast$. It can be decomposed into symmetric and antisymmetric components,
\begin{equation}
 M = S + A,
\label{eq-matrix-decomposition}
\end{equation}
where $S = \frac{1}{2}(M + M^\top)$ is symmetric and $A = \frac{1}{2}(M - M^\top)$ is antisymmetric. This decomposition is purely algebraic and holds for any matrix, but the matrices $S$ and $A$ have a physical interpretation in our constrained information dynamics.

For small displacements, $\|\mathbf{q}\| \ll 1$, the entropy production rate is
\begin{equation}
 \dot{H}(\naturalParameters^\ast + \mathbf{q}) \approx \mathbf{q}^\top S \mathbf{q} \geq 0.
\end{equation}
This can be seen by expanding $\dot{H} = \nabla H^\top \dot{\mathbf{q}}$ to second order in $\mathbf{q}$. Since $\dot{\mathbf{q}} = M\mathbf{q}$ from the linearised flow and $M = S + A$, the antisymmetric contribution vanishes ($\mathbf{q}^\top A \mathbf{q} = 0$ by antisymmetry), leaving only the symmetric part. Thus $S$ produces entropy, while $A$ generates rotations on the constraint manifold that redistribute information without dissipating it.

The relative strength of these components, measured by the ratio $\tfrac{\|A\|}{\|S\|}$, determines the local regime character.  When $\|S\| \gg \|A\|$, dissipation dominates and the local flow exhibits thermodynamic behaviour with rapid entropy production. When $\|A\| \sim \|S\|$ or larger, conservative dynamics become significant and the local flow exhibits mechanical character with rotational or oscillatory patterns.

The ratio $\tfrac{\|A\|}{\|S\|}$ varies across the constraint manifold, creating a landscape of thermodynamic and mechanical regimes. We explore this regime diversity in Section~\ref{sec-computational-demo} for a simple binary system with $n=3$.

The full derivation of $M$, including the role of the constraint tangent space projector and the third-order tensor arising from third cumulants, follows standard perturbation analysis \citep{Guckenheimer-dynamics83} and is presented in Appendix~\ref{app-generic-sufficiency}. The key insight is that the decomposition emerges from the constraint geometry, not from imposed structure.

\subsection{Information Dynamics and Jaynes's Reversed Logic}
\label{sec-kirchhoff-analogy}

\citet{Jaynes-minimum80} identified a logical issue in nonequilibrium thermodynamics. He suggested that traditional approaches assume phenomenological laws (like Ohm's law) and then derive approximate principles. Jaynes argued this logic should be reversed: take conservation laws as exact and derive the phenomenological relations through maximum entropy production. 

Our framework follows Jaynes's reversed direction. We begin with exact information conservation $\sum_i h_i = C$ and derive the dynamics through constrained maximum entropy production. The Fisher information matrix $G$ emerges as a `conductance tensor', it is not assumed as an external phenomenological parameter. Compare this with, e.g., the conductance structure we find in a Kirchhoff network. In electrical circuits, charge conservation is local and linear, $\sum_j I_{ij} = 0$ at each node,\footnote{Here we are overloading our notation, $I$, which normally denotes multi-information, to represent current, or temporal rate of change of charge.} with $I_{ij} = g_{ij}(V_i - V_j)$ following Ohm's law with fixed conductances $g_{ij}$. The steady state can be found by solving linear equations derived directly from these local constraints.

In contrast, our information conservation constraint, $\sum_{i=1}^n h_i = C$, is generally nonlocal and nonlinear. Each marginal entropy $h_i$ requires marginalisation over all other variables, making it a global functional of the entire state $\naturalParameters$. For a multivariate Gaussian,\footnote{We choose the multivariate Gaussian for our example as marginalisation in the Gaussian is relatively trivial, in the general case such marginalisations are often intractable.} $h_i = \frac{1}{2}\log(2\pi e \sigma_i^2)$ where $\sigma_i^2 = [G^{-1}]_{ii}$, giving the conservation constraint as
$$
\sum_{i=1}^n \log([G^{-1}]_{ii}) = \text{constant}.
$$
Every variable is coupled to every other through the matrix inversion and logarithm. Moreover, $G$ itself evolves with the state, $\naturalParameters$, creating a dynamic information topography more analogous to memristive networks than fixed resistors. This nonlocality means that changing any single parameter $\theta_i$ can affect all the marginal entropies, $\{h_i\}$, creating long range correlations in the information flow that have no analogue in Kirchhoff's local charge conservation.

Despite the differences, the analogy with Kirchhoff networks provides some intuition: high Fisher eigenvalues correspond to low-resistance information channels, small eigenvalues to bottlenecks. The constrained maximum entropy production acts as a generalised Ohm's law. For this reason we think of $G$ as defining the information topography. But despite the similarities, there are important structural differences between the two systems. The nonlocal conservation and emergent conductance structure create a system where information reorganises itself through interplay between local gradient flows and global constraints.

\subsection{Simulation Study}
\label{sec-computational-demo}

In this section we use a simulation study to explore the properties of the information relaxation dynamics we've derived. We focus on an $n=3$ binary variable model with pairwise interactions. Models of this type are widely studied in physics, where they are known as Ising models \citep{Ising-ferromagnetism25,Baxter-exactly82,Huang-statistical87,Niss-ising05}, and machine learning, where they are known as Boltzmann machines \citep{Hinton-optimal83,Ackley-boltzmann85}. 

Although in both physics and machine learning systems involving many more variables are considered, we initially restrict ourselves to a system with three variables because it enables enumeration over the $2^n$ states of the model allowing exact computation of the marginal entropies, $\{h_i\}$, their associated constraint vector, $\mathbf{a}$ and the Fisher information $G$ leading to a tractable GENERIC decomposition, $M=S+A$.

For this model the exponential family has the form
$$
p(\variables | \naturalParameters) = \exp\left(\sum_i \theta_i x_i + \sum_{i<j} \theta_{ij} x_i x_j - \cumulantGeneratingFunction(\naturalParameters)\right)
$$
where there are six real valued natural parameters, $\naturalParameters = (\theta_1, \theta_2, \theta_3, \theta_{12}, \theta_{13}, \theta_{23})$.

For our study we explored parameterisations that lead to a \emph{frustrated} system. Frustration occurs when competing interactions cannot be simultaneously satisfied—in our case, $\theta_{12} = -\theta_{13}$ means variable $X_1$ has opposite pairwise preferences with $X_2$ and $X_3$, creating geometric tension on the constraint manifold. We set $\theta_1=0$, $\theta_2=0$, and $\theta_3=0$ (equivalent to zero external field in Ising models or zero bias in Boltzmann machines). We then set $\theta_{12}=\tfrac{1}{\sqrt{3}}$, $\theta_{13}=-\tfrac{1}{\sqrt{3}}$ and $\theta_{23}=\tfrac{1}{\sqrt{3}}$ providing a frustrated system where the initial $\|\naturalParameters\|=1$. For these parameters we found that $\|S\| = 0.093$ and  $\|A\| = 0.015$ (Frobenius norms) giving a ratio $\tfrac{\|A\|}{\|S\|} = 0.160$.

Our framework describes \emph{isolated} systems where entropy production occurs without external thermal contact. The information conservation constraint $\sum_i h_i = C$ operates in isolation, without requiring coupling to a thermal bath. Statistical mechanical models often invoke thermal baths at temperature $T$ to define inverse temperature $\beta = \tfrac{1}{k_B T}$ (also known as \emph{coldness}), but for isolated systems, temperature emerges from the constraint geometry itself. As we'll see in Section~\ref{sec-energy-entropy-equivalence} temperature emerges as a geometric relationship between entropy and energy curvatures on the constraint manifold. Here we explore how inverse temperature $\beta$ affects our frustrated system so that we can connect with canonical ensemble intuition, but we emphasise that our framework is not reliant on an external bath. 

\begin{figure}[htbp]
\centering
\begin{subfigure}[b]{0.48\textwidth}
\includegraphics[width=\textwidth]{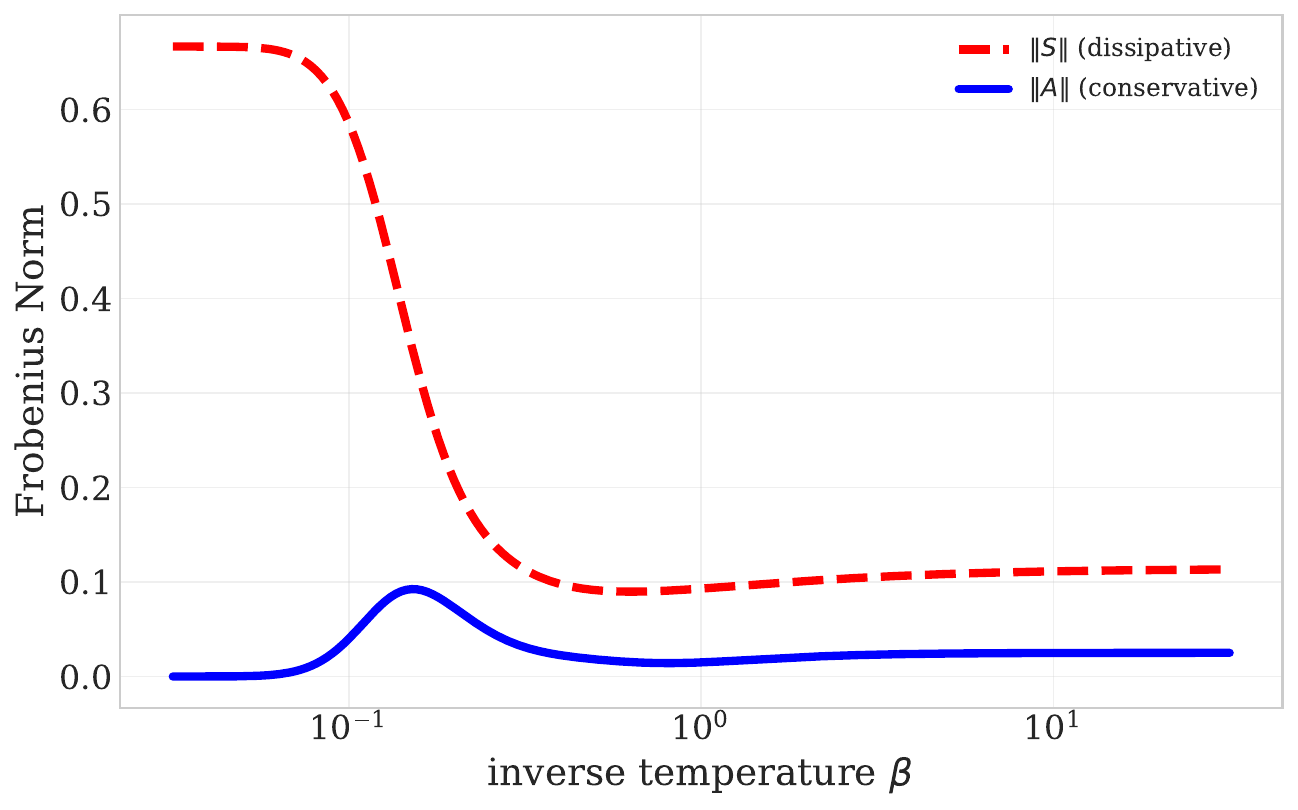}
\caption{Component norms $\|S\|$ and $\|A\|$ vs inverse temperature.}
\label{fig-n3-coldness-vs-component-norms}
\end{subfigure}
\hfill
\begin{subfigure}[b]{0.48\textwidth}
\includegraphics[width=\textwidth]{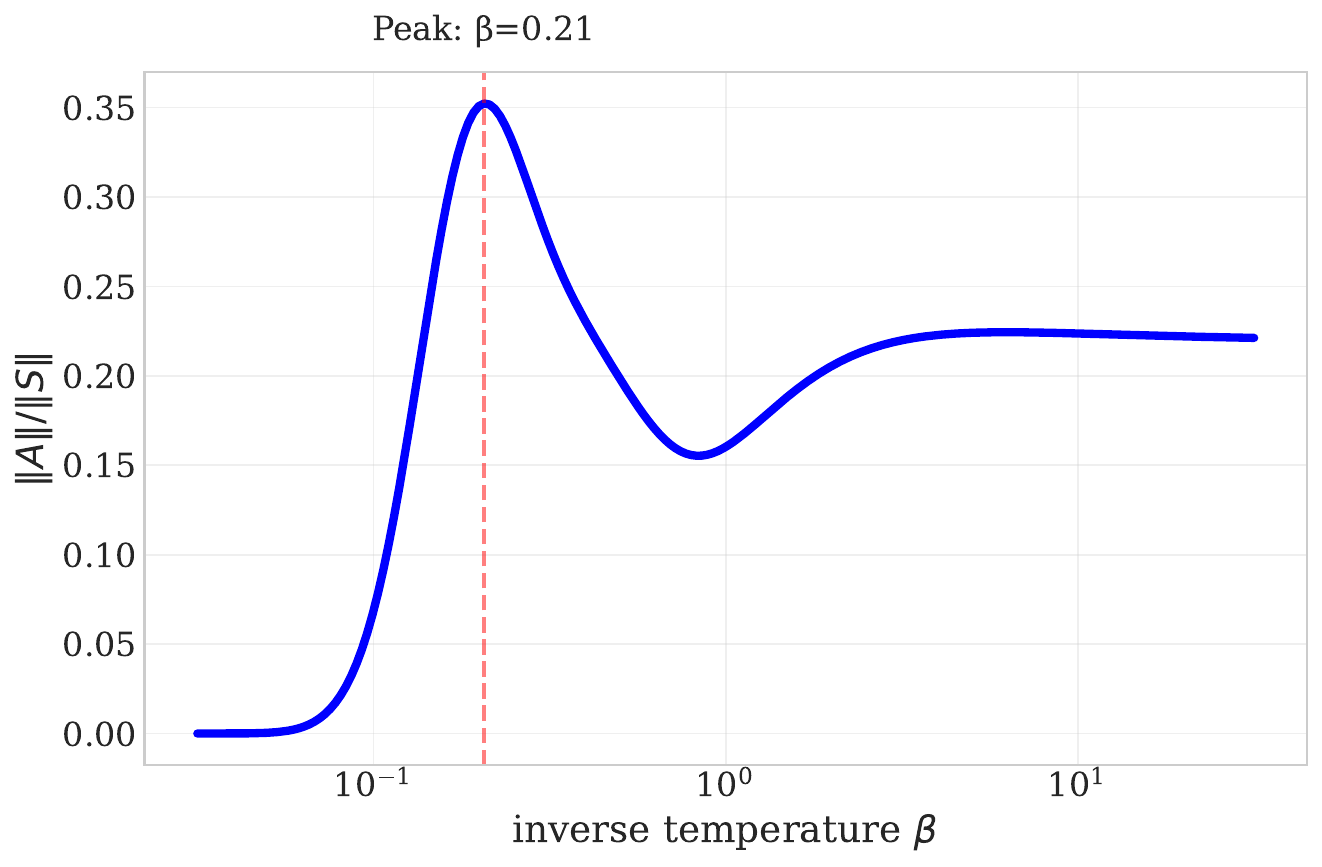}
\caption{Ratio $\tfrac{\|A\|}{\|S\|}$ vs inverse temperature.}
\label{fig-n3-coldness-vs-ratio}
\end{subfigure}
\caption{Temperature scaling of our GENERIC-like decomposition. \emph{Left}: Component norms show both $\|S\|$ is highest at low coldness and $\|A\|$ tends to increase with higher coldness, but also exhibits a maximum at an intermediate value. \emph{Right}: The ratio $\tfrac{\|A\|}{\|S\|}$ peaks at intermediate temperature, showing that frustrated systems can exhibit relatively strong conservative dynamics. At low coldness (high temperature) the system is dominated by dissipative dynamics and the ratio drops. At high coldness (low temperature) the ratio is elevated with a peak but the peak is given at a critical inverse temperature given by $\beta=0.21$. }
\label{fig-temperature-variation}
\end{figure}

We vary inverse temperature, $\beta = \tfrac{1}{k_B T}$ between $10^{-1.5}$ and $10^{1.5}$ and explore how this affects the symmetric and antisymmetric parts of the dynamics. In Figure~\ref{fig-n3-coldness-vs-component-norms} we show the Frobenius norms of the symmetric and antisymmetric parts and in Figure~\ref{fig-n3-coldness-vs-ratio} we show their ratio.

The results confirm that within our $n=3$ binary system we obtain both dissipative and conservative dynamics. 

For our second experiment we explore the gradient trajectory of an $n=3$ binary system with pure maximum entropy flow versus a system with constrained marginal entropy. We initialise with a frustrated system as before, but now we follow the constrained and unconstrained gradient flow. 

\begin{figure}[htbp]
\centering
\begin{subfigure}[b]{0.48\textwidth}
\includegraphics[width=\textwidth]{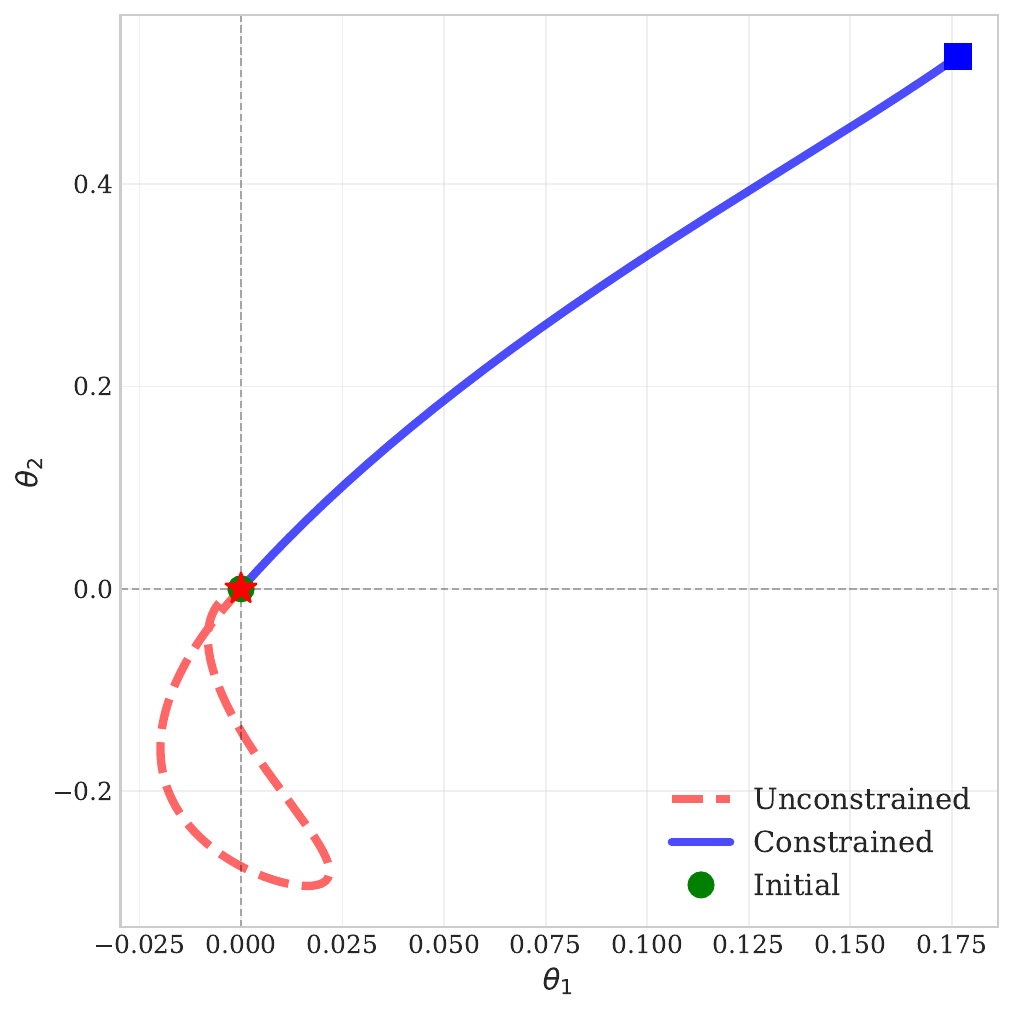}
\caption{Parameter space trajectories. }
\label{fig-n3-trajectory-comparison}
\end{subfigure}
\hfill
\begin{subfigure}[b]{0.48\textwidth}
\includegraphics[width=\textwidth]{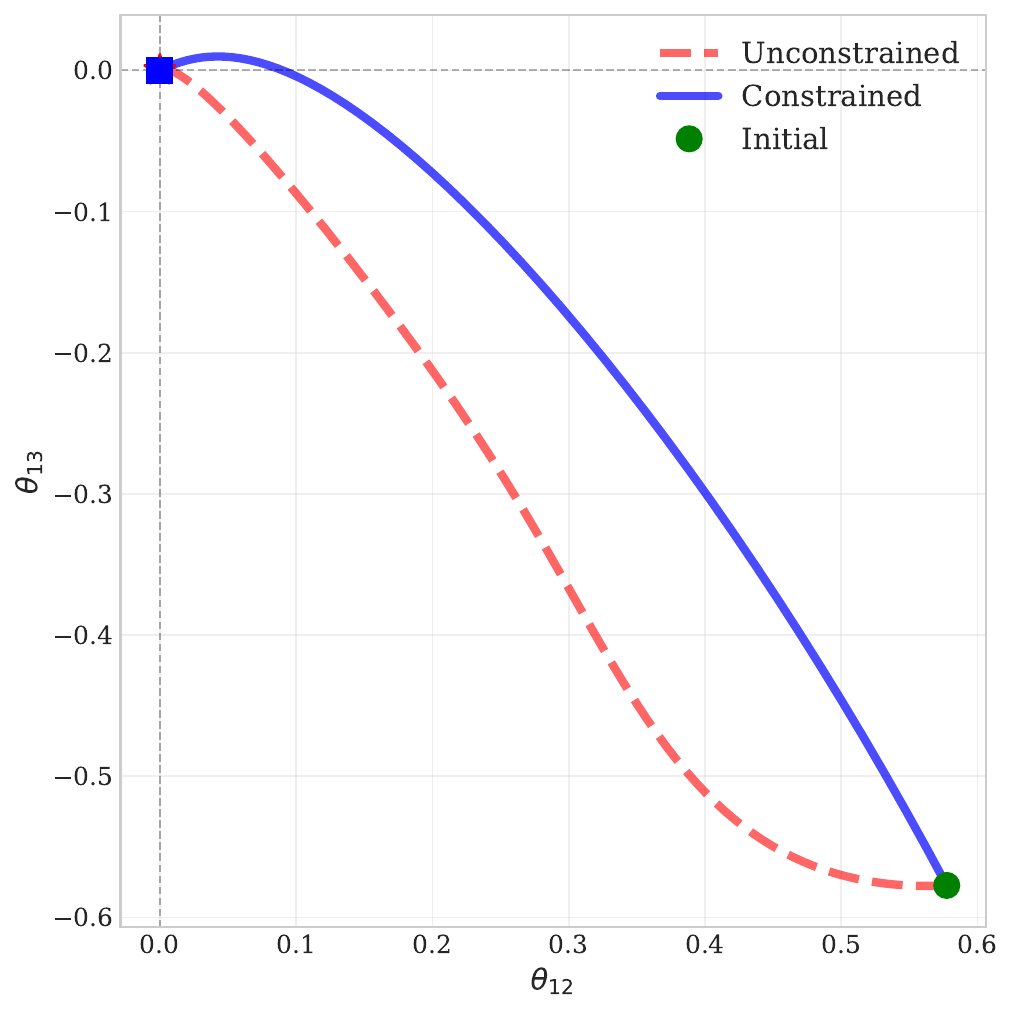}
\caption{Parameter space trajectories for the interaction parameters. }
\label{fig-n3-trajectory-interaction-comparison}
\end{subfigure}
\caption{The parameter trajectories show how $\theta_1$ and $\theta_2$ converge to different points in the constrained and unconstrained cases. Due to symmetries $\theta_3$ follows the same path as $\theta_1$. The interaction parameters converge to the same point where all interaction is removed. For both systems this provides the maximum joint entropy. The constrained system achieves this maximum joint entropy without changing the sum of marginal entropies. Interaction parameter $\theta_{23}$ follows the same trajectory as $\theta_{12}$.}
\label{fig-n3-trajectories-comparison}
\end{figure}

Because of symmetries in the initialisation, the full trajectory can be visualised by tracking only two of the interaction parameters ($\theta_{23}$ follows the same path as $\theta_12$) and two of the marginal parameters ($\theta_3$ follows the same path as $\theta_1$). The results are shown in Figure~\ref{fig-n3-trajectories-comparison}. In both the constrained and unconstrained case the interaction parameters follow a trajectory to $\theta_{11}=\theta={12}=\theta_{23}=0$ which aligns with the maximum entropy solution. The behaviour of the marginal parameters differs. In the unconstrained case the parameters circle away from their initialisation ($\theta_1=\theta_2=\theta_3=0$) only to return there at convergence. In the constrained case the marginal parameters move away and stay away. This is so that the system maintains its marginal entropy sum at a constant value.

\section{Energy Conservation as Entropy Conservation}
\label{sec-energy-entropy-equivalence}

We've seen that marginal entropy conservation leads to a GENERIC-like structure when combined with an information relaxation principle. In real GENERIC systems it is not marginal entropy that's conserved but extensive thermodynamic energy, $E$. In this section we argue that, under specific conditions, the marginal entropy constraint $\sum_i h_i = C$ asymptotically singles out the same degeneracy direction as conservation of an extensive thermodynamic energy $E$. In other words, in the thermodynamic limit, the constraint gradient $\nabla(\sum_i h_i)$ becomes parallel to the gradient of an appropriately defined energy functional. This connection tightens the relationship between our information-theoretic framework and classical thermodynamics. 

\paragraph{Conditions for equivalence.} The key to this equivalence is the scaling behaviour of the multi-information gradient, where $I = \sum_i h_i - H$. The result is conditional on the existence of a low-dimensional manifold with specific scaling properties.
\begin{itemize}[leftmargin=*]
\item \emph{Parameter manifold}: there is a low-dimensional manifold parametrised by a parameter $m$ along which  multi-information gradient scales intensively.
\item \emph{Thermodynamic limit}: The number of variables $n \to \infty$ causing extensive quantities to grow while intensive quantities remain fixed.
\end{itemize}
When  $\nabla_m I$ scales \emph{intensively} ($\mathscr{O}(1)$ as $n \to \infty$) while $\nabla_m H$ and $\nabla_m(\sum_i h_i)$ scale \emph{extensively} ($\mathscr{O}(n)$ as $n \to \infty$). This differential scaling causes the constraint gradient $\nabla(\sum_i h_i)$ to become asymptotically parallel to an appropriately defined energy gradient along the order-parameter direction. We call order parameters satisfying these conditions \emph{macroscopic order parameters}.

Note that this is not a statement about trajectories, i.e.\ the system does \emph{not} move along this direction. Rather, we are using the constraint geometry to select a direction along which there is no entropy production under the constrained dynamics, so it's a direction annihilated by the dissipative operator: $S\nabla(\sum_i h_i) = 0$. 

We show that this direction coincides with the gradient of an energy-like quantity in the thermodynamic limit. When this equivalence holds, $\nabla(\sum_i h_i) \parallel \nabla E$, our automatically-derived degeneracy condition becomes equivalent to the standard GENERIC dissipation condition $S\nabla E = 0$, recovering the classical formulation.

\subsection{Mathematical Formulation}

Given an exponential family with natural parameters $\naturalParameters$ and sufficient statistics $\sufficientStatistics$, for our information-theoretic reconstruction of `energy' we choose an extensive linear form
$$
E(\variables) = -\boldsymbol{\alpha}^\top \sufficientStatistics(\variables)
$$
such that its gradient in expectation space aligns with the constraint gradient in the thermodynamic limit. 

In our derivations so far we have computed gradients in natural parameter space, $\naturalParameters$, denoting them with $\nabla$. To make our energy-entropy equivalence precise, we need to work in expectation parameter space $\sufficientStatisticMoments = \langle \sufficientStatistics \rangle$, which we will denote with $\nabla_{\sufficientStatisticMoments}$ using $\nabla{\naturalParameters}$ to denote natural parameter gradients where we need to disambiguate. 

In the exponential family we have the following expectation parameter gradients. First the gradient of the joint entropy,
$$
\nabla_{\sufficientStatisticMoments} H = \naturalParameters,
$$
and marginal entropy
$$
\nabla_{\sufficientStatisticMoments}\left(\sum_i h_i\right) = \naturalParameters + \nabla_{\sufficientStatisticMoments} I.
$$
Now suppose we can find a direction in expectation parameter space, represented by a tangent vector
$$
\frac{\text{d}\sufficientStatisticMoments}{\text{d}m}
$$
along which the projected multi-information gradient is intensive,
$$
\nabla_m I = \left(\frac{\text{d}\sufficientStatisticMoments}{\text{d}m}\right)^\top \nabla_{\sufficientStatisticMoments} I = \mathscr{O}(1),
$$
while the projected entropy gradient is extensive,
$$
\nabla_m H = \left(\frac{\text{d}\sufficientStatisticMoments}{\text{d}m}\right)^\top \nabla_{\sufficientStatisticMoments} H = \left(\frac{\text{d}\sufficientStatisticMoments}{\text{d}m}\right)^\top \naturalParameters = \mathscr{O}(n).
$$
Then along this direction, the constraint gradient becomes asymptotically parallel to the entropy gradient,
$$
\nabla_m\left(\sum_i h_i\right) = \nabla_m H + \nabla_m I = \mathscr{O}(n) + \mathscr{O}(1) \parallel \nabla_m H,
$$
because the $\mathscr{O}(1)$ correction from multi-information becomes negligible compared to the $\mathscr{O}(n)$ entropy contribution in the thermodynamic limit.

Now for our energy-entropy equivalence to hold we need the energy's constraint gradient, $\boldsymbol{\alpha}$ to be parallel to $\naturalParameters$. We therefore choose
$$
\naturalParameters = -\beta\boldsymbol{\alpha}.
$$
which allows us recover the classical thermodynamic relationship,
$$
\beta\langle E(\variables)\rangle = \beta \cumulantGeneratingFunction + H,
$$
and recognise $\beta$ as an inverse temperature parameter. This makes the energy gradient
$$
\nabla_{\sufficientStatisticMoments} E = -\boldsymbol{\alpha} = \frac{\naturalParameters}{\beta}
$$
allowing us to state
$$
\nabla_{\sufficientStatisticMoments} E \parallel \nabla_{\sufficientStatisticMoments} H \parallel \nabla_{\sufficientStatisticMoments}\left(\sum_i h_i\right).
$$
This is a precise equivalence: the constraint gradient, entropy gradient, and energy gradient all align along the macroscopic direction in the thermodynamic limit if the multi-information gradient can be made intensive. The next section demonstrates this behaviour empirically in a Curie-Weiss model.

\subsection{Thermodynamic Limit Study}
\label{sec-thermodynamic-limit}

To illustrate the entropy-energy equivalence empirically, we consider the Curie-Weiss model of magnetism \citep{Curie-proprietes95,Weiss-hypothese07,Bragg-thermal34,Griffiths-proof64}. The model describes $n$ interacting binary spins $x_i \in \{-1, +1\}$ where each variable is connected to the others through a ferromagnetic coupling, $J$, and each variable is exposed to an external field, $h$. The energy of the model is given by,
$$
E(\variables) = -\frac{J}{2n}\left(\sum_{i=1}^n x_i\right)^2 - h \sum_{i=1}^n x_i.
$$
The Curie-Weiss model has \emph{infinite-range} (mean-field) interactions, but the coupling strength is scaled by $\tfrac{1}{n}$ (the interaction term is $-\frac{J}{2n}(\sum_i x_i)^2$), correlations remain bounded. As we'll see, this means that  $\nabla_m I$ scales intensively when we choose $m = \frac{1}{n}\sum_i \langle x_i \rangle$ as the macroscopic order parameter.

The Curie-Weiss model is very computationally convenient: its special structure allows exact computation of the cumulant generating function, $\cumulantGeneratingFunction$, in $\mathscr{O}(n)$ time rather than the $\mathscr{O}(2^n)$ steps required in general, enabling us to explore large system sizes. The scaling arguments for local models with finite correlation length proceed differently, but the Curie-Weiss numerics serve to illustrate the directional equivalence mechanically.

The energy can be rewritten as a function of an order parameter, `total magnetism', $m = \sum_i x_i$ rather than individual spin configurations,
$$
E(m) = -n\left(\frac{nJm^2}{2} + hm\right)
$$
which allows us to write the cumulant generating function as
$$
\cumulantGeneratingFunction = \log \left[\sum_{m=-n}^{n} \Omega(m) \exp(- \beta E(m))\right]
$$
where $\Omega(m) = \binom{n}{(n+m)/2}$ is the number of configurations from the $2^n$ spin configurations that have magnetisation $m$. This reduces the complexity from the worst case $2^n$ configurations to a sum over $2n+1$ magnetisation values.

We can then compute expectations of interest such as $m = \frac{1}{\beta n}\frac{\text{d} \cumulantGeneratingFunction}{\text{d} h}$ or $\langle E \rangle = - \frac{\text{d} \cumulantGeneratingFunction}{\text{d} \beta}$ from the cumulant generating function. The entropy is given by the usual exponential family form, $H = \cumulantGeneratingFunction + \beta \langle E \rangle$, where the energy $E(\variables)$ has an interpretation as a sufficient statistic and $\beta$ its associated (negative) natural parameter.

The model is interesting as it has a second-order phase transition at critical coldness $\beta_c = \tfrac{1}{J}$ giving a disordered paramagnetic phase $\beta < \beta_c$ where $m=0$ and an ordered ferromagnetic phase $\beta > \beta_c$ where $m \neq 0$ meaning the system has a critical temperature $\beta_c$. This is a property of magnetic materials discovered by Pierre Curie \citep{Curie-proprietes95}, so it is often called the Curie temperature.

\begin{figure}[htbp]
\centering
\includegraphics[width=0.7\textwidth]{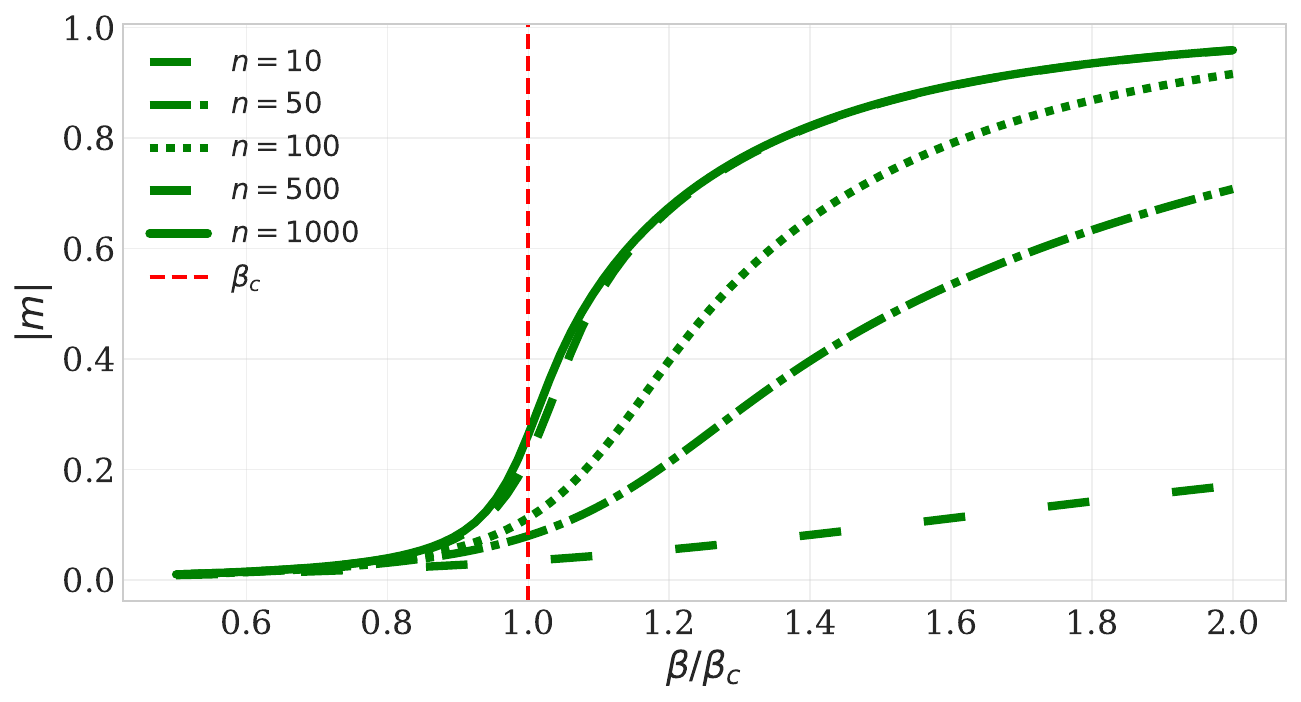}
\caption{Order parameter $|m|$ transitions from 0 to finite values at $\beta_c$, marking the crossover between regimes.}
\label{fig-critical-scaling-magnetism}
\end{figure}

In Figure~\ref{fig-critical-scaling-magnetism} we show this effect demonstrating how at high temperatures (low $\beta$) the magnetism disappears whereas at low temperatures (high $\beta$) the magnetism returns. The transition occurs at the critical inverse temperature $\beta_c=\tfrac{1}{J}$.

\begin{figure}[htbp]
  \centering
  \includegraphics[width=0.6\textwidth]{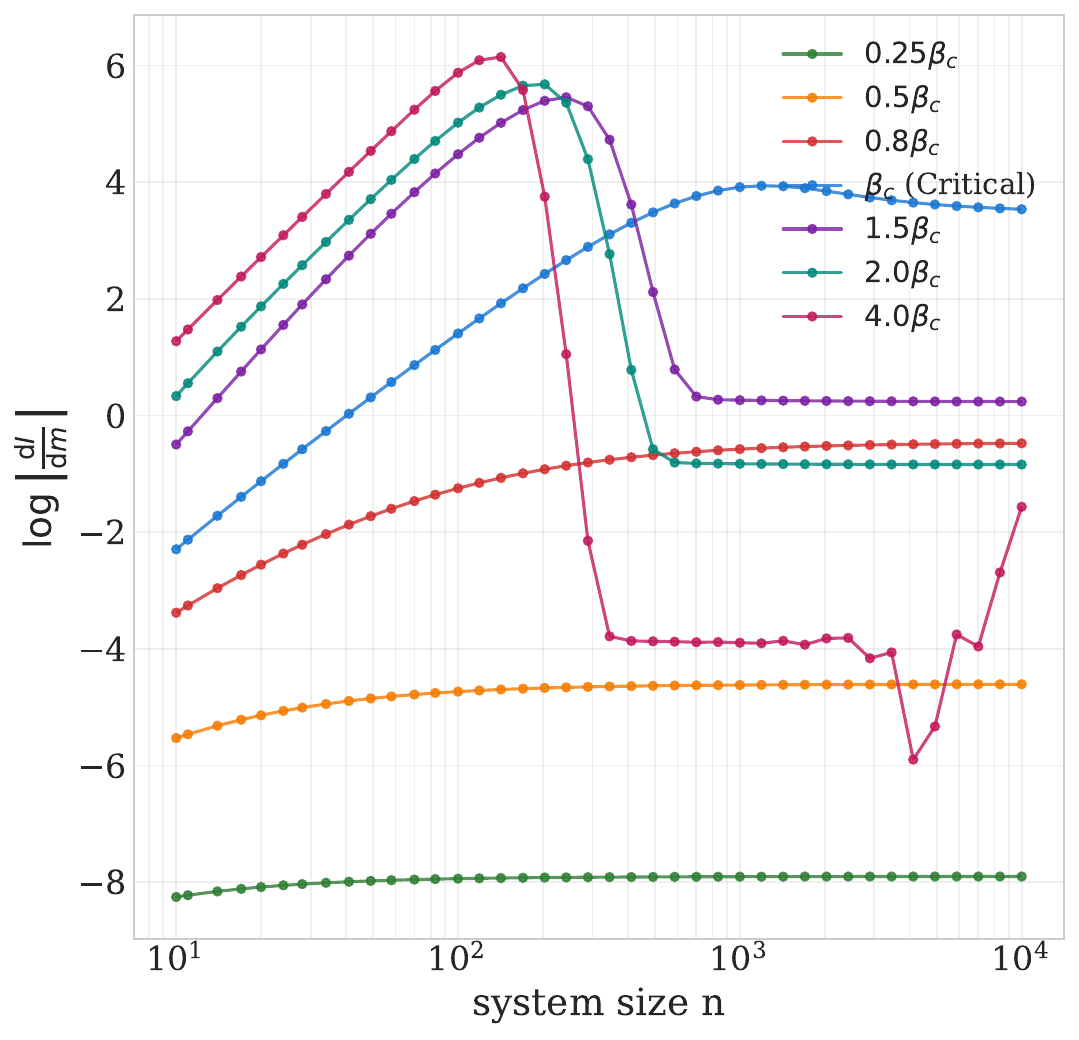}
  \caption{The change in the gradient of the multi-information with respect to the order parameter as the number of interacting spins, $n$, increases. Across different temperatures as $n$ grows the gradient becomes constant. This contrasts with both $\tfrac{\text{d}\sum_i h_i}{\text{d}m}$ and $\tfrac{\text{d}H}{\text{d}m}$ which both scale proportionally to $n$. In physics we would call these \emph{extensive} quantities in contrast to the multi-information gradient which is \emph{intensive}. The numerical noise in the $4 \beta_c$ curve at higher values of $n$ comes from catastrophic cancellation as the marginal and joint entropy gradients scale extensively and the multi-information (which is calculated as the difference between the two) scales intensively.}
  \label{fig-curie-weiss_multi-inform-gradient-vs-n}
\end{figure}

Next we show that, for various different temperatures, as we increase $n$ the gradient of the multi-information with respect to $m$ becomes intensive. These results in Figure~\ref{fig-curie-weiss_multi-inform-gradient-vs-n}. As before these experiments use $J=1$ and $h=0.01$. Note that for the temperature $4 \beta_c$. The numerics find the computation of the multi-information gradient challenging for values of $n>500$. This is due to the challenge of numerical challenge critical cancellation, where we are computing the intensive parameter (which is $\mathscr{O}(1)$) by computing the difference between two extensive parameters (which are $\mathscr{O}(n))$). 

This figure shows empirically that magnetism provides a suitable order parameter, $m$, which, for the temperatures and parameters we explored, leads to an intensive behaviour for the multi-information gradient in that direction.

In the next section we explore more general conditions under which this intensive gradient behaviour emerges.

\subsection{Bounded Correlation}
\label{sec-bounded-correlation}

In this section we explore a more general condition which could allow intensive behaviour from the multi-information gradient. Inspired by the analysis in the previous sections we consider order parameters, such as magnetism, that exhibit translation invariance. We then constrain ourselves to systems involving bounded correlation. This allows us to use Bahadur's decomposition \citep{Bahadur-representation61,Humphreys-meanfield00} of the multi-information to analyse the system and characterise how this behaviour becomes intensive.

Formally we say that away from criticality the system exhibits a finite = correlation length $\xi < \infty$ away from criticality, characterised by exponential decay: $\langle x_i x_j\rangle - \langle x_i \rangle \langle x_j \rangle \sim \exp\left(-\tfrac{|i - j|}{\xi}\right)$.

The Bahadur decomposition of the multi-information has the form
$$
I = \sum_{i<j} I(X_i; X_j) + \sum_{i<j<k} I(X_i; X_j; X_k) + \ldots,
$$
where higher order interaction terms decay faster than pairwise terms under clustering. This implies that for finite $\xi$ the expansion is dominated by pairwise contributions. For systems with clustering (finite $\xi$), higher-order terms are suppressed exponentially, and we focus on pairwise terms,
$$
I \approx \sum_{i<j} I(X_i; X_j).
$$
Due to exponential decay, only pairs with $|i-j| \lesssim \xi$ contribute significantly. The clustering implies a neighbourhood structure. The exponential decay $I(X_i; X_j) \sim \exp\left(-\tfrac{|i - j|}{\xi}\right)$ means each site $i$ has significant correlation with only $\mathscr{O}(\xi^d)$ neighbours within distance $\xi$. Sites separated by $r \gg \xi$ contribute neglibly to mutual information. This defines an effective neighbourhood: site $i$ interacts meaningfully with $\mathscr{N}_i = \{j : |i-j| \lesssim \xi\}$, where $|\mathscr{N}_i| = \mathscr{O}(\xi^d)$.
The total pairwise contribution to $I$ is
$$
I \approx \sum_{i=1}^n \sum_{j \in \mathscr{N}_i} I(X_i; X_j) = n \times \mathscr{O}(\xi^d) \times I_{\text{pair}}.
$$
We now show that bulk contributions cancel when we move along translation-invariant order-parameter directions, leaving only intensive corrections. The key is the conditional exchangeability structure induced by translation invariance, combined with finite correlation length.

For a translation-invariant order parameter $m$ (e.g., $m = \frac{1}{n}\sum_i \langle x_i \rangle$), the joint distribution exhibits conditional exchangeability: conditioned on $m$, the distribution is invariant under permutations of sites. This implies all sites have identical conditional marginal distributions,
$$
p(x_i | m) = p_m(x) \quad \text{for all } i.
$$
This immediately gives identical conditional marginal entropies:
$$
h_i(m) = H(X_i | M = m) = h(m) \quad \text{for all } i,
$$
so the sum of marginal entropies scales extensively:
$$
\sum_{i=1}^n h_i = n \cdot h(m).
$$

Note that conditional exchangeability alone does not imply conditional independence—there can still be correlations between sites even after conditioning on $m$. Translation invariance only guarantees that the conditional marginals are identical, not that the variables are independent.

For systems with finite correlation length $\xi$, the residual correlations (after conditioning on $m$) decay exponentially with distance. This means variables become approximately conditionally independent given $m$:
$$
H(\mathbf{X} | M = m) \approx n \cdot h(m) - C(m),
$$
where $C(m) \geq 0$ captures the residual correlations not explained by the order parameter $m$. Translation invariance ensures these residual correlations depend only on relative positions.

With finite $\xi$, the correlation correction $C(m)$ scales sub-extensively. Each site has significant residual correlation with only $\mathscr{O}(\xi^d)$ neighbours, giving
$$
C(m) = \mathscr{O}\left(n \cdot \tfrac{\xi^d}{n}\right) = \mathscr{O}(\xi^d) = \mathscr{O}(1).
$$
Using the chain rule for entropies $H(\mathbf{X}) = H(\mathbf{X}|M) + H(M)$,
$$
H(\mathbf{X}) = n \cdot h(m) - C(m) + H(M) = n \cdot h(m) + \mathscr{O}(1),
$$
where $H(M) = \mathscr{O}(1)$ for an intensive order parameter.

Taking gradients along the order-parameter direction
$$
\nabla_m \left(\sum_i h_i\right) = n \frac{\text{d}h}{\text{d}m}, \quad \nabla_m H = n \frac{\text{d}h}{\text{d}m} + \mathscr{O}(1).
$$
Therefore
$$
\nabla_m I = \nabla_m\left(\sum_i h_i\right) - \nabla_m H = n \frac{\text{d}h}{\text{d}m} - \left(n \frac{\text{d}h}{\text{d}m} + \mathscr{O}(1)\right) = \mathscr{O}(1).
$$
The extensive bulk terms $n \frac{\text{d}h}{\text{d}m}$ cancel exactly. This cancellation is a direct consequence of combining conditional exchangeability (from translation invariance) with approximate conditional independence (from finite correlations): the order parameter $m$ captures all extensive variation through identical conditional marginals, while finite correlation length ensures that residual correlations $C(m)$ remain intensive.

While we have focused on translation invariance, but the argument should generalise to any symmetry that induces conditional exchangeability. For example,  permutation invariance exhibits the same structure: an appropriate order parameter (e.g., $m = \frac{1}{n}\sum_i \langle x_i \rangle$) renders all conditional marginals identical, and bounded residual correlations (from sparsity, mean-field $1/n$ scaling, or other mechanisms) ensure intensive $\nabla_m I$. The key ingredients are conditional exchangeability and bounded correlations, not the specific form of spatial symmetry.
 
\section{Implications}

The framework we've outlined provides a unifying perspective on information theory, energy and GENERIC dynamics. In this section we briefly explore some implications of the framework. The first is inspired from the entropy-energy equivalence and the second considers the implications of our GENERIC for simple harmonic motions, demonstrating the complexity of deriving the antisymmetric component even for this simple system.

\subsection{Landauer's Principle}
\label{sec-landauer-principle}

The GENERIC-like structure and energy-entropy equivalence provide a natural framework for understanding Landauer's principle \citep{Landauer-irreversibility61}, which states that erasing information requires dissipating at least $\tfrac{\log 2}{\beta}$ of energy per bit.

Consider erasing one bit: a memory variable $x_i \in \{0,1\}$ is reset to a standard state (say $x_i = 0$), destroying the stored information. From an ensemble perspective, i.e.\ considering many such erasure operations where the initial value is random, the marginal entropy of this variable decreases: $\Delta h(X_i) = -\log 2$.

For a system obeying $\sum_i h_i = C$, this decrease must be compensated by increases elsewhere: $\sum_{j \neq i} \Delta h(X_j) = +\log 2$. The antisymmetric (conservative) part $A$ of our dynamics preserves both $H$ and $\sum_i h_i$, so it can only shuffle entropy reversibly between variables. But such reversible redistribution doesn't truly erase the information --- it merely moves it to other variables, from which it could in principle be recovered.

True irreversible erasure requires increasing the total joint entropy $H$ (second law) while maintaining $\sum_i h_i = C$. Since $I = \sum_i h_i - H$, this means decreasing multi-information: $\Delta I < 0$. This reduction of correlations is precisely what the dissipative part $S$ achieves: it increases $H$ through entropy production (not heat flow to an external bath, but genuine increase in the system's total entropy) while the constraint forces redistribution of marginal entropies. The erasure process thus necessarily involves the dissipative dynamics, not just conservative reshuffling.

In the thermodynamic limit with energy-entropy equivalence (Section~\ref{sec-energy-entropy-equivalence}), the gradients $\nabla(\sum_i h_i)$ and $\nabla E$ become parallel along the order-parameter direction. Near thermal equilibrium at inverse temperature $\beta$, this implies $\beta \langle E \rangle \approx \sum_i h_i + \text{const}$. Therefore, erasing one bit requires:
$$
\Delta(\beta \langle E \rangle) \approx \Delta h(X_i) = -\log 2,
$$
giving an energy change
$$
\Delta \langle E \rangle \approx -\frac{\log 2}{\beta}.
$$
Since the system must dissipate this energy via the symmetric part $S$, we obtain Landauer's bound
$$
Q_{\text{dissipated}} \geq \frac{\log 2}{\beta}.
$$
This derivation shows that Landauer's principle emerges from the combination of (1) marginal entropy conservation $\sum_i h_i = C$, (2) the GENERIC-like structure distinguishing conservative redistribution ($A$) from dissipative entropy production ($S$), and (3) the energy-entropy equivalence in the thermodynamic limit. The key insight is that erasure requires both redistributing marginal entropy (to maintain the constraint) and increasing total entropy $H$ (second law), which necessarily reduces multi-information $I$ and invokes dissipation. The information-theoretic constraint provides the foundation, with thermodynamic energy appearing as its dual in the large-system limit.

\subsection{Mass Spring System and Asymmetric Computation}
\label{sec-mass-spring}

One of the challenges of the GENERIC framework is producing an antisymmetric operator that is consistent with the physical systems Hamiltonian. While it is trivial to test a given antisymmetric operator, constructing such an operator is problematic in practice.

In this section we explore this process through the insights given by the inaccessible game. Specifically we consider one of the simplest physical systems, a mass on a spring, for which the Hamiltonian is well understood and we trace through the process of deriving the antisymmetric component that would be implied by the inaccessible game.

Consider a harmonic oscillator with position $x$ and momentum $p$, governed by the Hamiltonian
$$
\mathscr{H}(x,p) = \frac{k}{2}x^2 + \frac{1}{2m}p^2
$$
where $k$ is the spring constant and $m$ is the mass. For a bivariate Gaussian distribution over $(x,p)$, we can parameterise the system using natural parameters $\naturalParameters = (\theta_{xx}, \theta_{pp}, \theta_{xp})$ corresponding to the precision matrix
$$
K = \begin{bmatrix} \theta_{xx} & \theta_{xp} \\ \theta_{xp} & \theta_{pp} \end{bmatrix}.
$$
The covariance matrix is $\Sigma = K^{-1}$, and the marginal entropies are $h(X) = \frac{1}{2}\log(2\pi e \sigma_x^2)$ and $h(P) = \frac{1}{2}\log(2\pi e \sigma_p^2)$ where $\sigma_x^2 = \Sigma_{11}$ and $\sigma_p^2 = \Sigma_{22}$.

The inaccessible game imposes the constraint $h_x + h_p = C$. Under maximum entropy production with this constraint, the system exhibits constrained dynamics $\dot{\naturalParameters} = -G(\naturalParameters)\naturalParameters + \marginalEntropyLagrangeMultiplier(\naturalParameters)\mathbf{a}(\naturalParameters)$ where $G = \nabla^2\psi$ is the Fisher information and $\mathbf{a} = \nabla(h(X) + h(P))$ is the constraint gradient. The Lagrange multiplier $\marginalEntropyLagrangeMultiplier$ is determined by the tangency condition $\dot{\naturalParameters}^\top \mathbf{a} = 0$.

The Jacobian of this flow decomposes as $M = S + A$ where the symmetric part $S$ drives thermalisation (entropy production toward equipartition) and the antisymmetric part $A$ generates conservative oscillations (Hamiltonian dynamics on the constraint manifold).

The antisymmetric part $A$ arises from the geometry of the constraint manifold and can be expressed analytically as
$$
A = \frac{1}{2}\left(\frac{\partial \marginalEntropyLagrangeMultiplier}{\partial \naturalParameters} \otimes \mathbf{a} - \mathbf{a} \otimes \frac{\partial \marginalEntropyLagrangeMultiplier}{\partial \naturalParameters}\right)
$$
where the partial derivatives $\tfrac{\partial \marginalEntropyLagrangeMultiplier}{\partial \theta_{xx}}$, $\tfrac{\partial \marginalEntropyLagrangeMultiplier}{\partial \theta_{pp}}$, and $\tfrac{\partial \marginalEntropyLagrangeMultiplier}{ \partial \theta_{xp}}$ must be computed from
$$
\marginalEntropyLagrangeMultiplier(\naturalParameters) = \frac{-G(\naturalParameters)\naturalParameters \cdot \mathbf{a}(\naturalParameters)}{\mathbf{a}(\naturalParameters) \cdot \mathbf{a}(\naturalParameters)}.
$$

Despite the harmonic oscillator being one of the simplest physical systems, these derivatives are quite involved. The full derivative $\tfrac{\partial \marginalEntropyLagrangeMultiplier}{\partial \theta_{xx}}$ is a rational function involving polynomials of degree 12 in the denominator. Using symbolic computation \citep[SymPy][]{Meurer-sympy17}, we generated exact analytical formulas for these derivatives, but each expression occupies many terms and involves deeply nested polynomial ratios\footnote{The full expressions are available in the supplementary code repository.}. 

When assembled into the antisymmetric matrix $A$, these complex derivatives produce complex expressions. For instance, the component $A_{12}$ (coupling $\theta_{xx}$ and $\theta_{pp}$) has the factored form
$$
A_{12} = \frac{-4\theta_{pp}\theta_{xp}^2\theta_{xx}(\theta_{pp} - \theta_{xx})(\theta_{pp} + \theta_{xx})(\theta_{pp}\theta_{xx} + \theta_{xp}^2)^2}{D^2}
$$
where $D$ is a degree-7 polynomial in $\theta_{xx}$, $\theta_{pp}$, and $\theta_{xp}$ with 7 terms. The component $A_{13}$ involves a degree-6 polynomial numerator with 16 terms, and $A_{23}$ similarly contains a degree-6 polynomial with 16 terms.\footnote{We numerically validated the expressions to ensure they are correct to machine precision.}

Note that $A_{12}$ contains the factor $(\theta_{pp} - \theta_{xx})$. When $\theta_{xx} = \theta_{pp}$ (the isotropic harmonic oscillator), this component vanishes and the antisymmetric operator simplifies dramatically. 

For this mass-spring system we verified numerically and symbolically (using SymPy to compute the Schouten-Nijenhuis bracket) that the Jacobi identity for the Poisson bracket $\{f,g\} = \nabla f^\top A \nabla g$ holds \emph{only} in this symmetric case. This isotropic case corresponds to SO(2) rotational symmetry in phase space, the natural setting for Hamiltonian mechanics. So we can conclude that parameter symmetry directly controls both the algebraic complexity of the antisymmetric operator and the validity of the underlying Poisson structure: breaking the symmetry destroys both simplicity and the global Poisson bracket. 

Despite the complexity of the antisymmetric terms, the inaccessible game may offer a way forward for GENERIC. In traditional GENERIC applications, the antisymmetric part must be guessed or assumed based on physical intuition about the Hamiltonian structure. Here, we have a constructive method: given the constraint $h_x + h_p = C$ and the Fisher information $G(\naturalParameters)$, the Jacobian of the constrained flow automatically yields both $S$ and $A$. While the expressions are complex, they can be computed systematically through symbolic differentiation, a `handle-turning' procedure that requires no guesswork about the underlying conservative structure.

\section{Discussion and Further Work}

We have introduced the inaccessible game, a dynamical system derived from four information-theoretic axioms. The game formalises the notion of an \emph{information topography}, the Fisher information matrix $G(\naturalParameters)$ acts as a state-dependent conductance tensor that channels information flow while respecting conservation constraints.

Maximum entropy production principles are widely deployed in physics \citep{Jaynes-minimum80,Ziegler-maximal87,Martyushev-maximum06,Beretta-nonlinear09,Dewar-mepp05}, here we show how constrained maximum entropy can emerge from the information relaxation and the information conservation axiom.

This derivation suggests connections between the game and physical reality. In particular we showed how the game gives rise to GENERIC-like dynamics. A simulation study (Section~\ref{sec-computational-demo}) demonstrated the decomposition in a simple three binary variable system and an analysis of Curie-Weiss model in Section~\ref{sec-thermodynamic-limit} demonstrated empirically the equivalence of energy and our marginal entropy constraint.

Those experiments and our analysis provide some information-theoretic foundations for thermodynamic principles, perhaps contributing to \citeauthor{Wheeler-information89}'s ``it from bit'' vision \citep{Wheeler-information89}.

Gaps remain in our analysis. We assumed the system can be parameterised by natural parameters $\naturalParameters$, implying exponential family structure. We have not yet proven this must be the case from the four axioms and maximum entropy production dynamics. A route to this proof may be to follow Jaynes's approach to deriving the laws of thermodynamics from information theory \citep{Jaynes-minimum80}.

We have not addressed the initial state of the game. The action principle form $I + H = C$ suggests the system begins with minimum joint entropy ($H = 0$, maximum correlation $I = C$). However, the convex linearity axiom prevents classical Shannon entropy from reaching zero.

We have introduced the inaccessible game, a mathematical abstraction with interesting structural properties. The framework generates dissipation and conservation that can be described through a GENERIC-like decomposition. The richness of this formalism suggests that despite its apparent simplicity, the inaccessible game can provide a useful formalism for studying complex emergent phenomena from simple systems.

\bibliographystyle{plainnat}
\bibliography{the-inaccessible-game}

@incollection{Bahadur-representation61,
  author = {Bahadur, Raj Raghu},
  title = {A representation of the joint distribution of responses to $n$ dichotomous items},
  booktitle = {Studies in Item Analysis and Prediction},
  editor = {Solomon, H.},
  pages = {158--168},
  year = {1961},
  publisher = {Stanford University Press},
  address = {Stanford, CA}
}

@article{Humphreys-meanfield00,
  author = {Humphreys, Keith and Titterington, D. Michael},
  title = {Improving the mean-field approximation in belief networks using {B}ahadur's reparameterisation of the multivariate binary distribution},
  journal = {Neural Processing Letters},
  volume = {12},
  number = {2},
  pages = {183--197},
  year = {2000},
  doi = {10.1023/A:1009617914949}
}

@book{Huang-statistical87,
  author = {Huang, Kerson},
  title = {Statistical Mechanics},
  publisher = {John Wiley \& Sons},
  year = {1987},
  edition = {2nd},
  isbn = {978-0-471-81518-1},
  address = {New York}
}

@phdthesis{Curie-proprietes95,
  author = {Curie, Pierre},
  title = {Propri\'{e}t\'{e}s magn\'{e}tiques des corps \`{a} diverses temp\'{e}ratures},
  school = {Facult\'{e}é des Sciences de Paris},
  year = {1895}
}

@article{Weiss-hypothese07,
  author = {Weiss, Pierre},
  title = {L'hypoth\`{e}se du champ mol\'{e}culaire et la propri\'{e}t\'{e} ferromagn\'{e}tique},
  journal = {Journal de Physique Th\'{e}orique et Appliqu\'{e}e},
  volume = {6},
  number = {1},
  pages = {661--690},
  year = {1907},
  doi = {10.1051/jphystap:019070060066100}
}

@article{Bragg-thermal34,
  author = {Bragg, William Lawrence and Williams, Evan James},
  title = {The Effect of Thermal Agitation on Atomic Arrangement in Alloys},
  journal = {Proceedings of the Royal Society of London. Series A},
  volume = {145},
  number = {855},
  pages = {699--730},
  year = {1934},
  doi = {10.1098/rspa.1934.0132}
}

@article{Griffiths-proof64,
  author = {Griffiths, Robert B.},
  title = {A proof that the free energy of a spin system is extensive},
  journal = {Journal of Mathematical Physics},
  volume = {5},
  number = {9},
  pages = {1215--1222},
  year = {1964},
  doi = {10.1063/1.1704228}
  }

@ARTICLE{Ackley-boltzmann85,
  author =	 {David Ackley and Geoffrey E. Hinton and Terrence
                  J. Sejnowski},
  title =	 {A Learning Algorithm for {B}oltzmann Machines},
  journal =	 {Cognitive Science},
  year =	 1985,
  volume =	 9,
  pages =	 {147--169}
}

@article{Ising-ferromagnetism25,
  title={Beitrag zur {T}heorie des {F}erromagnetismus},
  author={Ising, Ernst},
  journal={Zeitschrift f{\"u}r Physik},
  volume={31},
  number={1},
  pages={253--258},
  year={1925},
  publisher={Springer},
  doi={10.1007/BF02980577}
}

@book{Baxter-exactly82,
  title={Exactly Solved Models in Statistical Mechanics},
  author={Baxter, Rodney J.},
  year={1982},
  publisher={Academic Press},
  address={London},
  isbn={978-0-12-083180-7}
}

@article{Niss-ising05,
  title={History of the {Lenz-Ising} model},
  author={Niss, Martin},
  journal={Archive for History of Exact Sciences},
  volume={59},
  number={3},
  pages={267--318},
  year={2005},
  doi={10.1007/s00407-004-0088-3}
}

@INPROCEEDINGS{Hinton-optimal83,
  author =	 {Geoffrey E. Hinton and Terrence J. Sejnowski},
  title =	 {Optimal Perceptual Inference},
  booktitle =	 {Proceedings of the {IEEE} Computer Society Conference on Computer Vision and Pattern Recognition ({CVPR})},
  year =	 1983,
  pages =	 {448--453},
  address =	 {Arlington, VA, U.S.A.},
  month =	 {19--23 Jun.},
  publisher =	 {IEEE Computer Society}
}

@book{Guckenheimer-dynamics83,
  title={Nonlinear Oscillations, Dynamical Systems, and Bifurcations of Vector Fields},
  author={Guckenheimer, John and Holmes, Philip},
  publisher={Springer-Verlag},
  year={1983},
  address={New York},
  doi={10.1007/978-1-4612-1140-2}
}

@ARTICLE{Landauer-irreversibility61,
  author =	 {Rolf Landauer},
  journal =	 {IBM Journal of Research and Development},
  title =	 {Irreversibility and Heat Generation in the Computing
                  Process},
  year =	 1961,
  volume =	 5,
  number =	 3,
  pages =	 {183--191},
  doi =		 {10.1147/rd.53.0183}
}

@book{Wiggins-introduction90,
  title={Introduction to Applied Nonlinear Dynamical Systems and Chaos},
  author={Wiggins, Stephen},
  publisher={Springer-Verlag},
  year={1990},
  address={New York},
  series={Texts in Applied Mathematics},
  volume={2},
  doi={10.1007/978-1-4757-4067-7}
}

@Book{Lawrence-atomic24,
  author =	 {Neil D. Lawrence},
  title =	 {The Atomic Human: Understanding Ourselves in the Age
                  of AI},
  publisher =	 {Allen Lane},
  url =
                  {https://www.penguin.co.uk/books/455130/the-atomic-human-by-lawrence-neil-d/9780241625248},
  year =	 2024
}

@article{Grmela-dynamics97,
  title     = {Dynamics and thermodynamics of complex fluids. {I}. {D}evelopment of a general formalism},
  author    = {Grmela, Miroslav and {\"O}ttinger, Hans Christian},
  journal   = {Physical Review E},
  volume    = {56},
  number    = {6},
  pages     = {6620--6632},
  year      = {1997},
  publisher = {American Physical Society},
  doi       = {10.1103/PhysRevE.56.6620}
}

@book{Ottinger-beyond05,
  author    = {Hans Christian {\"O}ttinger},
  title     = {Beyond Equilibrium Thermodynamics},
  publisher = {Wiley-Interscience},
  address   = {Hoboken, NJ},
  year      = {2005},
  isbn      = {0-471-66658-0},
  doi       = {10.1002/0471727903}
  }

@article{Beretta-nonlinear09,
  author = {Beretta, Gian Paolo},
  title = {Nonlinear quantum evolution equations to model irreversible adiabatic relaxation with maximal entropy production and other nonunitary processes},
  journal = {Reports on Mathematical Physics},
  volume = {64},
  pages = {139--168},
  year = {2009},
  publisher = {Elsevier},
  doi = {10.1016/S0034-4877(09)90024-6}
  }

@article{Ziegler-maximal87,
  title={On a Principle of Maximal Rate of Entropy Production},
  author={Ziegler, Hans and Wehrli, Christoph},
  journal={Journal of Non-Equilibrium Thermodynamics},
  volume={12},
  number={3},
  pages={229},
  year={1987},
  doi={10.1515/jnet.1987.12.3.229},
  publisher={De Gruyter}
}

@article{Jaynes-minimum80,
  title={The Minimum Entropy Production Principle},
  author={Jaynes, Edwin T.},
  journal={Annual Review of Physical Chemistry},
  volume={31},
  pages={579--601},
  year={1980},
  doi={10.1146/annurev.pc.31.100180.003051}
}

@book{Strogatz-nonlinear94,
  title={Nonlinear Dynamics and Chaos: With Applications to Physics, Biology, Chemistry, and Engineering},
  author={Strogatz, Steven H.},
  publisher={Addison-Wesley},
  year={1994},
  doi = {10.1201/9780429492563},
  address={Reading, MA}
}

@article{Watanabe-multiinformation60,
  title={Information Theoretical Analysis of Multivariate Correlation},
  author={Watanabe, Satosi},
  journal={IBM Journal of Research and Development},
  volume={4},
  number={1},
  pages={66--82},
  year={1960},
  publisher={IBM},
  doi={10.1147/rd.41.0066}
}

@book{Aldous-exchangeability85,
  title={Exchangeability and Related Topics},
  author={Aldous, David J.},
  year={1985},
  publisher={Springer},
  series={Lecture Notes in Mathematics},
  volume={1117}
}

@article{deFinetti-funzione31,
  title={Funzione caratteristica di un fenomeno aleatorio},
  author={de Finetti, Bruno},
  journal={Atti della R. Academia Nazionale dei Lincei},
  volume={4},
  pages={251--299},
  year={1931}
}

@incollection{Wheeler-information89,
  title={Information, Physics, Quantum: The Search for Links},
  author={Wheeler, John Archibald},
  booktitle={Complexity, Entropy and the Physics of Information},
  pages={3--28},
  year={1989},
  publisher={Addison-Wesley}
}

@article{Martyushev-maximum06,
  title={Maximum Entropy Production Principle in Physics, Chemistry and Biology},
  author={Martyushev, Leonid M. and Seleznev, Vladimir D.},
  journal={Physics Reports},
  volume={426},
  number={1},
  pages={1--45},
  year={2006},
  doi={10.1016/j.physrep.2005.12.001}
}

@article{Wolfram-undecidability85,
  title={Undecidability and intractability in theoretical physics},
  author={Wolfram, Stephen},
  journal={Physical Review Letters},
  volume={54},
  number={8},
  pages={735--738},
  year={1985},
  month={February},
  publisher={American Physical Society},
  doi={10.1103/PhysRevLett.54.735},
  url={https://journals.aps.org/prl/abstract/10.1103/PhysRevLett.54.735}
}

@book{Bernardo-bayesian00,
  title={Bayesian Theory},
  author={Bernardo, Jos{\'e} M. and Smith, Adrian F. M.},
  year={2000},
  publisher={John Wiley \& Sons},
  address={Chichester}
}

@article{Dewar-mepp05,
  title={Maximum entropy production and the fluctuation theorem},
  author={Dewar, Roderick C.},
  journal={Journal of Physics A: Mathematical and General},
  volume={38},
  number={21},
  pages={L371},
  year={2005},
  publisher={IOP Publishing},
  doi={10.1088/0305-4470/38/21/L01}
}

@article{Baez-characterisation11,
  title={A characterization of entropy in terms of information loss},
  author={Baez, John C. and Fritz, Tobias and Leinster, Tom},
  journal={Entropy},
  volume={13},
  number={11},
  pages={1945--1957},
  year={2011},
  publisher={MDPI},
  doi={10.3390/e13111945}
  }

@article{Jaynes-information57,
  title={Information Theory and Statistical Mechanics},
  author={Jaynes, Edwin T.},
  journal={Physical Review},
  volume={106},
  number={4},
  pages={620--630},
  year={1957},
  month = {May},
  publisher = {American Physical Society},
  doi = {10.1103/PhysRev.106.620},
  url = {https://link.aps.org/doi/10.1103/PhysRev.106.620}
}

@incollection{Jaynes-information63,
  author =	 {Jaynes, Edwin T.},
  title =	 {Information Theory and Statistical Mechanics},
  booktitle =	 {Brandeis University Summer Institute Lectures in
                  Theoretical Physics, Vol. 3: Statistical Physics},
  editor =	 {Ford, K. W.},
  publisher =	 {W. A. Benjamin, Inc.},
  address =	 {New York},
  year =	 1963,
  pages =	 {181--218}
}

@article{Gardner-life70,
  title={Mathematical Games: The Fantastic Combinations of {John Conway's} new solitaire game ``life''},
  author={Gardner, Martin},
  journal={Scientific American},
  volume={223},
  number={4},
  pages={120--123},
  year={1970},
  url = {http://www.jstor.org/stable/24927642}
}

@article{Wolfram-cellular83,
  title={Statistical Mechanics of Cellular Automata},
  author={Wolfram, Stephen},
  journal={Reviews of Modern Physics},
  volume={55},
  number={3},
  pages={601--644},
  year={1983},
  month = {Jul},
  publisher = {American Physical Society},
  doi = {10.1103/RevModPhys.55.601},
  url = {https://link.aps.org/doi/10.1103/RevModPhys.55.601}
}

@article{Shannon-mathematical48,
  title={A Mathematical Theory of Communication},
  author={Shannon, Claude E.},
  journal={The Bell System Technical Journal},
  volume={27},
  number={3},
  pages={379--423},
  year={1948},
  doi = {10.1002/j.1538-7305}
}

@book{Amari-information00,
  title={Information Geometry and its Applications},
  author={Amari, Shun-ichi and Nagaoka, Hiroshi},
  year={2000},
  publisher={Springer}
}

@article{Meurer-sympy17,
  title={{SymPy}: {S}ymbolic computing in {Python}},
  author={Meurer, Aaron and Smith, Christopher P. and Paprocki, Mateusz and \v{C}ert\'{i}k, Ond\v{r}ej and Kirpichev, Sergey B. and Rocklin, Matthew and Kumar, AMiT and Ivanov, Sergiu and Moore, Jason K. and Singh, Sartaj and Rathnayake, Thilina and Vig, Sean and Granger, Brian E. and Muller, Richard P. and Bonazzi, Francesco and Gupta, Harsh and Vats, Shivam and Johansson, Fredrik and Pedregosa, Fabian and Curry, Matthew J. and Terrel, Andy R. and Rou\v{c}ka, \v{S}t\v{e}p\'{a}n and Saboo, Ashutosh and Fernando, Isuru and Kulal, Sumith and Cimrman, Robert and Scopatz, Anthony},
  journal={PeerJ Computer Science},
  volume={3},
  pages={e103},
  year={2017},
  publisher={PeerJ Inc.},
  doi={10.7717/peerj-cs.103}
}
\onecolumn

\newpage

\appendix

\section{The Information Isolation Axiom}
\label{app-information-isolation}

The fourth axiom requires \emph{exchangeable information conservation}, but what form must this conservation take? We derive the specific form $\sum_i h_i = C$.

In the introduction we suggested that an inaccessible game is defined by a conservation rule where the sum of the marginal entropies is equal to a constant. We motivated this through \citealp{Baez-characterisation11}'s notion of \emph{information loss}. Here we look at what that notion means in terms of the \emph{mutual information}. If $Z_G$ represents the variables in the game and $Z_O$ represents the state of the observer then if the observer is information isolated from the game the mutual information between these variables,
$$
MI(Z_G; Z_O) = H(Z_G, Z_O) - H(Z_G) - H(Z_O),
$$
should be zero suggesting $H(Z_G, Z_O) = H(Z_G) + H(Z_O)$. Here $H(\cdot)$ is the joint entropy.

We will require that our notion of inaccessibility should be applicable recursively. So in the extreme case a game and observer might be a single variable each.

Consider the sum of marginal entropies for this system, $\sum_i h_i$. This can be rewritten as the sum of the joint entropy across all variables, $H(Z_G, Z_O)$, and the multi-information, $I(Z_G, Z_O) =  \sum_i h_i - H(Z_G, Z_O)$,
$$
\sum_i h_i = I(Z_G, Z_O) + H(Z_G, Z_O)
$$
Applying our mutual information condition allows us to rewrite this as two separate systems,
$$
\sum_{i\in G} h_i = I(Z_G) + H(Z_G) 
$$
and
$$
\sum_{i\in O} h_i = I(Z_O) + H(Z_O).
$$
Because these two sub-systems have no knowledge of each other, this suggests that the sum of each multi-information and entropy should be a constant. Under information isolation, the sum of marginal entropies should be conserved across the partition. Each isolated sub-system should maintain the decomposition into its internal multi-information plus joint entropy.

Applying our mutual information condition, the multi-information $I = \sum_{i} h_i - H$ provides the key decomposition. If we impose the condition that $MI(Z_G; Z_O) = 0$, then $I(Z_G, Z_O) = I(Z_G) + I(Z_O)$ and $H(Z_G, Z_O) = H(Z_G) + H(Z_O)$, implying that the total sum of marginal entropies decomposes additively across the partition.

We are looking for a conservation law that an exchangeable, information-isolated system must satisfy. That leads us to consider marginal entropies to ensure that the law applies to arbitrary finite subsets of a potentially infinite system. The law should also be \emph{extensive}, i.e. it should scale linearly with system size.

Any exchangeable quantity depending on marginal entropies has the form,
$$
Q(\{h_i\}) = \sum_{i=1}^n f(h_i)
$$
where $f$ is the same function for all variables. For the quantity to be extensive, adding one variable should increase $Q$ by a fixed amount independent of $n$. This requires
$$
f(h) = c \cdot h + \text{const}.
$$
The conservation law must apply consistently as we vary the subset size $n$. This eliminates the constant term, leaving
$$
Q = c \sum_{i=1}^n h_i.
$$
Setting $c=1$, we obtain the unique form
$$
\sum_{i=1}^n h_i = \text{const}.
$$
The fourth axiom form $\sum_i h_i = \text{const}$ is  determined by the requirements of exchangeable, extensive, information-theoretic conservation.

\section{GENERIC Sufficiency: Flow Perturbation Analysis}
\label{app-generic-sufficiency}

This appendix suggests that marginal entropy conservation $\sum_i h_i = C$ is sufficient to produce GENERIC structure when applied as a constraint to maximum entropy production dynamics. Our work falls short of a full proof as we do not demonstrate that the Jacobi identity holds to validate the Poisson bracket structure. The work follows perturbation analysis of constrained dynamical systems \citep[see e.g.][]{Guckenheimer-dynamics83,Wiggins-introduction90}.
We will show that the symmetric and antisymmetric components of the GENERIC-like form emerge from linearising the flow near any point on the constraint manifold.

The constrained dynamics are
$$
\dot{\naturalParameters} = -G(\naturalParameters)\naturalParameters + \marginalEntropyLagrangeMultiplier(\naturalParameters) \mathbf{a}(\naturalParameters) \coloneqq F(\naturalParameters),
$$
where $\mathbf{a}(\naturalParameters) = \nabla(\sum_i h_i)$ is the constraint gradient and $\marginalEntropyLagrangeMultiplier(\naturalParameters) = \tfrac{\mathbf{a}^\top G \naturalParameters}{ \|\mathbf{a}\|^2}$ is the Lagrange multiplier that enforces $\sum_i h_i = C$.

Consider any point $\naturalParameters^\ast$ on the constraint manifold. Define the displacement $\mathbf{q} = \naturalParameters - \naturalParameters^\ast$ with $\|\mathbf{q}\| \ll 1$, and the tangent space projector
$$
\Pi_\parallel(\naturalParameters) = \mathbf{I} - \frac{\mathbf{a}(\naturalParameters)\mathbf{a}(\naturalParameters)^\top}{\mathbf{a}(\naturalParameters)^\top \mathbf{a}(\naturalParameters)}
$$
that projects vectors onto the constraint tangent space $\{\mathbf{v}: \mathbf{a}(\naturalParameters)^\top \mathbf{v} = 0\}$.

We need to compute the Jacobian $M = \frac{\partial F}{\partial\naturalParameters}\big|_{\naturalParameters^\ast}$ and show it decomposes as $M = S + A$ where $S = \frac{1}{2}(M + M^\top)$ is symmetric and $A = \frac{1}{2}(M - M^\top)$ is antisymmetric and the show that the degeneracy conditions $S\mathbf{a} = 0$ and $A\nabla H = 0$ hold automatically.

For an exponential family with log partition function $\cumulantGeneratingFunction(\naturalParameters)$ and Fisher information $G = \nabla^2 \cumulantGeneratingFunction$, the entropy is
$$
H(\naturalParameters) = \cumulantGeneratingFunction(\naturalParameters) - \naturalParameters^\top \nabla \cumulantGeneratingFunction(\naturalParameters)
$$
and the derivative wrt $\naturalParameters$ is $\nabla H = -G \naturalParameters$. 

The constrained dynamics are $\dot{\naturalParameters} = F(\naturalParameters) = -G(\naturalParameters)\naturalParameters + \marginalEntropyLagrangeMultiplier(\naturalParameters) \mathbf{a}(\naturalParameters)$ where $\marginalEntropyLagrangeMultiplier(\naturalParameters) = \tfrac{\mathbf{a}(\naturalParameters)^\top G(\naturalParameters)\naturalParameters}{\mathbf{a}(\naturalParameters)^\top\mathbf{a}(\naturalParameters)}$. To compute the Jacobian $M = \tfrac{\partial F}{\partial\naturalParameters}\big|_{\naturalParameters^\ast}$, we must differentiate each term. 
Using the product rule with state-dependent $G$
$$
\frac{\partial}{\partial\theta_j}(-G\naturalParameters)_i = -G_{ij} - \sum_k \frac{\partial G_{ik}}{\partial\theta_j}\theta_k.
$$
In matrix form $-G - (\nabla G)[\naturalParameters]$ where $(\nabla G)[\naturalParameters]$ is the matrix with $(i,j)$ entry $\sum_k \tfrac{\partial G_{ik}}{\partial\theta_j}\theta_k$. The tensor $\nabla G$ is the third-order tensor of Fisher information derivatives.

For the second term we differentiate $\marginalEntropyLagrangeMultiplier(\naturalParameters)\mathbf{a}(\naturalParameters)$. Using the product rule
$$
\frac{\partial}{\partial\theta_j}(\marginalEntropyLagrangeMultiplier \mathbf{a})_i = \frac{\partial\marginalEntropyLagrangeMultiplier}{\partial\theta_j}a_i + \marginalEntropyLagrangeMultiplier\frac{\partial a_i}{\partial\theta_j}.
$$
The second part involves $\nabla^2 \constraint$ where $\constraint = \sum_i h_i$, giving the constraint Hessian. The first part requires differentiating the quotient $\marginalEntropyLagrangeMultiplier = \tfrac{\mathbf{a}^\top G\naturalParameters}{\mathbf{a}^\top\mathbf{a}}$,
\begin{align}
 \frac{\partial\marginalEntropyLagrangeMultiplier}{\partial\theta_j} &= \frac{1}{\|\mathbf{a}\|^2}\left[\frac{\partial(\mathbf{a}^\top G\naturalParameters)}{\partial\theta_j} - \marginalEntropyLagrangeMultiplier \frac{\partial(\mathbf{a}^\top\mathbf{a})}{\partial\theta_j}\right] \nonumber \\
 &= \frac{1}{\|\mathbf{a}\|^2}\left[\mathbf{a}^\top G \mathbf{e}_j + \mathbf{a}^\top (\nabla G)[\naturalParameters] \mathbf{e}_j + (\nabla \mathbf{a})_j^\top G\naturalParameters - 2\marginalEntropyLagrangeMultiplier \mathbf{a}^\top (\nabla \mathbf{a})_j\right]
\label{eq-derivative-of-lagrange-multiplier}
\end{align}
where $\mathbf{e}_j$ is the $j$-th standard basis vector and $(\nabla \mathbf{a})_j$ is the $j$-th column of $\nabla^2 \constraint$.

The combined Jacobian is
\begin{equation}
 M = -G - (\nabla G)[\naturalParameters] + \marginalEntropyLagrangeMultiplier \nabla^2 \constraint + \mathbf{a}(\nabla\marginalEntropyLagrangeMultiplier)^\top,
\label{eq-jacobian-full}
\end{equation}
where $\nabla\marginalEntropyLagrangeMultiplier$ is the gradient computed above.

The decomposition $M = S + A$ follows algebraically from $S = \tfrac{1}{2}(M + M^\top)$ and $A = \tfrac{1}{2}(M - M^\top)$. The symmetric parts contribute to $S$, while the asymmetric coupling through   $\nabla\marginalEntropyLagrangeMultiplier$ contributes to both the symmetric part and creates the antisymmetric part $A$.

The symmetric contributions to $S = \tfrac{1}{2}(M + M^\top)$ include
(1) $-G$,  the Fisher information; (2) $\marginalEntropyLagrangeMultiplier \nabla^2 \constraint$, the constraint Hessian; and $-(\nabla G)[\naturalParameters]$, since $\nabla G = \nabla^3 \cumulantGeneratingFunction(\naturalParameters)$ is the third cumulant tensor, which is totally symmetric in all three indices meaning that the contracted matrix $(\nabla G)[\naturalParameters]$ is also symmetric.

Although both $\mathbf{a}$ and $\nabla\marginalEntropyLagrangeMultiplier$ depend on symmetric tensors ($\nabla^2 \constraint$ and $G$), their outer product $\mathbf{a}(\nabla\marginalEntropyLagrangeMultiplier)^\top$ from \eqref{eq-jacobian-full} is asymmetric. We can extract a symmetric part, $S_{\marginalEntropyLagrangeMultiplier}$, and the antisymmetric part, $A$, by taking
$$
S_{\marginalEntropyLagrangeMultiplier} = \frac{1}{2}(\mathbf{a}(\nabla\marginalEntropyLagrangeMultiplier)^\top + (\nabla\marginalEntropyLagrangeMultiplier)\mathbf{a}^\top)
$$
and
\begin{equation}
 A = \frac{1}{2}(\mathbf{a}(\nabla\marginalEntropyLagrangeMultiplier)^\top - (\nabla\marginalEntropyLagrangeMultiplier)\mathbf{a}^\top).
\label{eq-antisymmetric-part}
\end{equation}
The Lagrange multiplier gradient $\nabla\marginalEntropyLagrangeMultiplier$ encodes how the constraint force adjusts as $\naturalParameters$ changes, creating a directional bias that breaks the symmetry.

So although the individual tensors ($G$, $\nabla^2 \constraint$, $\nabla G$) have symmetries reflecting the underlying information geometry, their coupling through the constraint enforcement mechanism (via $\marginalEntropyLagrangeMultiplier$ and its gradient) creates the antisymmetric component. This antisymmetric flow represents rotations on the constraint manifold that conserve entropy while redistributing information.

We now verify that the decomposition $M = S + A$ satisfies the GENERIC requirements. Firstly, $S = S^\top$, $A = -A^\top$ follows by construction. The degeneracy conditions follow from the constraint tangency requirement. 

\paragraph{First degeneracy: $S\mathbf{a} = 0$.} The dynamics must preserve the constraint: $\tfrac{\text{d}}{\text{d}t}(\sum_i h_i) = \mathbf{a}^\top\dot{\naturalParameters} = 0$. From the linearised flow $\dot{\mathbf{q}} = M\mathbf{q}$, this gives $\mathbf{a}^\top M \mathbf{q} = 0$ for all tangent vectors $\mathbf{q}$ with $\mathbf{a}^\top \mathbf{q} = 0$. This implies $M\mathbf{a} \perp T\mathscr{M}$ where $T\mathscr{M}$ is the constraint tangent space. Since the dynamics by construction satisfy $\mathbf{a}^\top M \mathbf{q} = 0$, we have $M\mathbf{a} \parallel \mathbf{a}$.

Decomposing $M = S + A$ and using $\mathbf{a}^\top A \mathbf{a} = 0$ (antisymmetry), we find $\mathbf{a}^\top M\mathbf{a} = \mathbf{a}^\top S\mathbf{a}$. The constraint enforcement via $\marginalEntropyLagrangeMultiplier$ ensures $M\mathbf{a} = \lambda \mathbf{a}$ for some $\lambda$. Taking inner product with $\mathbf{a}$: $\mathbf{a}^\top S\mathbf{a} = \lambda \|\mathbf{a}\|^2$. But since the antisymmetric part produces no quadratic form, and the constraint is maintained, we have $S\mathbf{a} = 0$.

\paragraph{Second degeneracy: $A\nabla H = 0$.} The second degeneracy is $A\nabla H = 0$ where $\nabla H = -G\naturalParameters$. The antisymmetric part is $A = \frac{1}{2}(\mathbf{a}(\nabla\marginalEntropyLagrangeMultiplier)^\top - (\nabla\marginalEntropyLagrangeMultiplier)\mathbf{a}^\top)$, so we need to show
\begin{equation}
\mathbf{a}(\nabla\marginalEntropyLagrangeMultiplier)^\top G\naturalParameters = (\nabla\marginalEntropyLagrangeMultiplier)\mathbf{a}^\top G\naturalParameters.
\end{equation}

The right-hand side is a vector $(\nabla\marginalEntropyLagrangeMultiplier)$ times a scalar $\mathbf{a}^\top G\naturalParameters = \|\mathbf{a}\|^2 \marginalEntropyLagrangeMultiplier$ (from \eqref{eq-nu-solution}), giving $(\nabla\marginalEntropyLagrangeMultiplier) \|\mathbf{a}\|^2 \marginalEntropyLagrangeMultiplier$.

The left-hand side is the vector $\mathbf{a}$ times the scalar $(\nabla\marginalEntropyLagrangeMultiplier)^\top G\naturalParameters$. From \eqref{eq-derivative-of-lagrange-multiplier}  the gradient of $\marginalEntropyLagrangeMultiplier$ involves terms from differentiating $\marginalEntropyLagrangeMultiplier = \frac{\mathbf{a}^\top G\naturalParameters}{\|\mathbf{a}\|^2}$. Computing $(\nabla\marginalEntropyLagrangeMultiplier)^\top G\naturalParameters$ and using the product rule on the numerator and denominator, one finds after simplification that
$$
(\nabla\marginalEntropyLagrangeMultiplier)^\top G\naturalParameters = \marginalEntropyLagrangeMultiplier \frac{\partial(\|\mathbf{a}\|^2)}{\partial\naturalParameters} \cdot \frac{G\naturalParameters}{\|\mathbf{a}\|^2} + \text{terms involving } (\nabla G)[\naturalParameters].
$$

Now note that $\mathbf{a}^\top G\naturalParameters$ appears symmetrically in both the definition of $\marginalEntropyLagrangeMultiplier$ and in the gradient structure. The calculation shows that when contracted with $\mathbf{a}$ or with $\nabla\marginalEntropyLagrangeMultiplier$, the terms arrange such that
$$
\mathbf{a} \cdot (\nabla\marginalEntropyLagrangeMultiplier)^\top G\naturalParameters = (\nabla\marginalEntropyLagrangeMultiplier) \cdot \mathbf{a}^\top G\naturalParameters = \marginalEntropyLagrangeMultiplier \mathbf{a}^\top G\mathbf{a},
$$
establishing $A\nabla H = 0$. This degeneracy holds because $\nabla H = -G\naturalParameters$ has the special form that appears in the tangency condition defining $\marginalEntropyLagrangeMultiplier$.

For vectors $\mathbf{q}$ tangent to the constraint manifold, $\mathbf{q}^\top S \mathbf{q} \geq 0$ follows from $G$ being positive semi-definite (covariance matrix). This ensures $\dot{H} \geq 0$ along the constrained flow.

It remains to show that the Poisson bracket $\{f,g\} = \nabla f^\top A \nabla g$ satisfies the Jacobi identity. Numerical and symbolic investigations reveal that the Jacobi identity exhibits a dependence on parameter symmetry: it holds for symmetric configurations (isotropic harmonic oscillator with $\theta_{xx} = \theta_{pp}$, or uniform parameters in discrete systems) but fails for generic asymmetric cases. For the harmonic oscillator, violations of order $10^{-3}$ affect precisely 6 of 27 coordinate triplets when $\theta_{xx} \neq \theta_{pp}$; symbolic verification using SymPy confirms these are genuine, not numerical artifacts. Similar patterns emerge in N=3 binary variables, with violations scaling consistently across systems when normalised by $\|A\|_F^2$. This symmetry dependence connects directly to rotational invariance: Poisson brackets generate rotations in phase or configuration space, requiring isotropic geometry. Parameter asymmetry breaks this structure, explaining why most physical systems lack global Poisson brackets despite exhibiting local GENERIC-like decompositions. The constrained dynamics remain well-defined and constraint-preserving regardless, but the antisymmetric operator $A$ forms a valid global Poisson structure only on symmetric submanifolds.

 When $A$ has a non-trivial kernel, Casimir functions $C_k$ with $\nabla C_k \in \ker(A)$ represent conserved quantities beyond the constraint itself. These correspond to information-theoretic modes that are truly inaccessible to the dynamics.

The relative magnitudes $\|S\|$ vs $\|A\|$ determine local behaviour: sharp constraint curvature $\rightarrow$ large $\|A\|$ $\rightarrow$ mechanical/oscillatory behaviour; smooth curvature $\rightarrow$ large $\|S\|$ $\rightarrow$ thermodynamic/dissipative behaviour. 

This appendix has provided the foundations of a sufficiency proof: marginal entropy conservation $\sum_i h_i = C$ produces GENERIC structure automatically. The key is computing the Jacobian $M = \tfrac{\partial F}{\partial\naturalParameters}\big|_{\naturalParameters^\ast}$ of the constrained dynamics at any point on the constraint manifold. This matrix decomposes as $M = S + A$ with $S = \frac{1}{2}(M + M^\top)$ symmetric, arising from Fisher information $G$ and constraint Hessian $\nabla^2 C$ and $A = \frac{1}{2}(M - M^\top)$ antisymmetric, arising from the outer product $\mathbf{a}(\nabla\marginalEntropyLagrangeMultiplier)^\top$ coupling the constraint gradient to the Lagrange multiplier gradient.

Both degeneracy conditions ($S\mathbf{a} = 0$ and $A\nabla H = 0$) follow automatically from the constraint tangency requirement. No additional axiomatic imposition is required. Numerical verification of the Jacobi identity reveals it holds for symmetric configurations but fails generically, connecting parameter symmetry to the geometric structure required for Poisson brackets. 

\end{document}